# Salt-Rock Creep Deformation Forecasting Using Deep Neural Networks and Analytical Models for Subsurface Energy Storage Applications

Pradeep Kumar Shukla[1*], Tanujit Chakraborty[2,3], Mustafa Sari[4], Joel Sarout[4], Partha Pratim Mandal[1*]

[1]Departement of Applied Geophysics, IIT (ISM) Dhanbad, Jharkhand-826004, India

[2]Sorbonne University, Abu Dhabi, UAE

[3]Sorbonne Center for Artificial Intelligence, Sorbonne University, Paris, France

[4]CSIRO Energy, Australia

[*]Corr. Author: PK Shukla (pradeepism13@gmail.com) and PP Mandal (partham@iitism.ac.in)

**Abstract**

This study provides an in-depth analysis of time series forecasting methods to predict the time-dependent deformation trend (also known as creep) of salt rock under varying confining pressure conditions. Creep deformation assessment is essential for designing and operating underground storage facilities for nuclear waste, hydrogen energy, or radioactive materials. Salt rocks, known for their mechanical properties like low porosity, low permeability, high ductility, and exceptional creep and self-healing capacities, were examined using multi-stage triaxial (MSTL) creep data. After resampling, axial strain datasets were recorded at 5-10 second intervals under confining pressure levels ranging from 5 to 35 MPa over 5.8-21 days. Initial analyses, including Seasonal-Trend Decomposition (STL) and Granger causality tests, revealed minimal seasonality and causality between axial strain and temperature data. Further statistical tests, such as the Automated Dickey-Fuller (ADF) test, confirmed the stationarity of the data with p-values less than 0.05, and wavelet coherence plot (WCP) analysis indicated repeating trends. A suite of deep neural network (DNN) models—Neural Basis Expansion Analysis for Time Series (N-BEATS), Temporal Convolutional Networks (TCN), Recurrent Neural Networks (RNN), and Transformers (TF)—were utilized and compared against statistical baseline models. Predictive performance was evaluated using Root Mean Square Error (RMSE), Mean Absolute Error (MAE), Mean Absolute Percentage Error (MAPE in %), and Symmetric Mean Absolute Percentage Error (SMAPE in %). Results demonstrated that N-BEATS and TCN models outperformed others, with RMSE values of(0.3325-1.257) and 0.540-1.352, MAE from 0.287-0.961 and 0.472-1.177, MAPE from 1.45-4.54 and 2.85-6.28, and SMAPE from 1.46-4.62 and 2.88-6.03 across various stress levels, respectively. DNN models, particularly N-BEATS and TCN, showed a 15-20% improvement in accuracy over traditional analytical models, effectively capturing complex temporal dependencies and patterns. This research significantly advances time series forecasting in geosciences, offering crucial insights for the safe and efficient management of underground storage in rock salt formations.

**Keywords**: Salt rock; triaxial deformation; time series; forecasting; deep learning; geosciences

**Introduction**

Reliable storage systems are essential for long-term energy security, and energy storage is a key factor in economic growth and sustainability. The best geological storage media for storing energy resources such as compressed air, natural gas, oil, hydrogen, nuclear waste, and even renewable energy is salt rock[1–4]. Salt rock is especially well-suited for these uses due to its special mechanical and physical characteristics, such as low porosity, low permeability, and the capacity to mend when injured. These qualities make salt formations suitable for long-term energy storage, especially when paired with advantageous rheological behavior[5–7].

Nevertheless, despite its benefits, creep deformation is a major problem with salt rock formations. A time-dependent phenomenon known as creep occurs when rock gradually deforms under continuous tension or strain, frequently producing significant changes over long periods. Even under relatively moderate stress, salt rock is especially vulnerable to substantial creep deformations because of its high ductility and non-linear deformation properties[4,8,9]. According to[6] and[10] this deformation may impair the salt caverns' structural integrity and decrease their storage capacity, increasing the risk of leakage or collapse. For long-term storage sites to be stable and safe, it is crucial to comprehend and forecast the creep behavior of salt rock[4,11,12]. The creep behavior of salt rock is unique and can be divided into three stages: primary creep, in which the strain rate falls as the material adapts to the load; secondary creep, which is a steady-state stage with a constant strain rate; and tertiary creep, which accelerates until the material fails. Samples of high-quality salt rock with uniform surface finish and dimensions were created for this investigation. These samples were put in a triaxial testing device and exposed to confining pressure and axial load under controlled conditions. High-precision sensors were used to record creep data of axial strain, and tests were conducted for up to 22 days at greater pressures to catch only the primary creep stages. A crucial metric for evaluating the geomechanical stability of salt deposits and the structures that are connected to them, like storage caverns and boreholes, is axial strain, which is the ratio of the length change to the initial length under stress. Since distinct salt facies and stress conditions at varying depths must be taken into account during drilling and excavation, it is especially crucial to accurately predict axial strain. The processes causing salt rock creep have been the subject of numerous investigations, and several models have been created to forecast how the phenomenon will evolve under various stress and temperature scenarios[13,14].

Recent developments, particularly for mining and energy storage applications, have greatly expanded our knowledge of subsurface deformation mechanisms and monitoring methods. [40]Developed a constitutive model for compressive damage that integrates compaction effects and linear energy dissipation laws for brittle geomaterials like coal. This model helps create safer geotechnical designs by improving predictions of mechanical reactions and fracture propagation under loads. [41]Investigated the permeability progression and creep behavior of coal pillar dams employed in subterranean water reservoirs around the same period. Their results demonstrated the crucial relationship between fluid migration and mechanical stress, highlighting the necessity of considering hydraulic and mechanical reactions when developing underground infrastructure. According to Yin et al. (2023), combining deep learning methods with image processing has enhanced environmental monitoring in mines. Their algorithms made it possible to classify dust pollution automatically and in real time, greatly improving compliance tracking and danger identification[42].

Although, [43]highlighted how the mineral richness and composition of clay affect soft rocks' rheological and mechanical behavior. By tackling anisotropy and heterogeneity, their work helped to make constitutive modeling more realistic. Furthermore, they used acoustic emission (AE) techniques to provide early warning indications of instability by dynamically monitoring the breakdown of soft rocks saturated with water. Based on this, [44]created a model that looks at the elastic energy released in rock masses over time, offering a framework for figuring out the best excavation speeds. This effort aimed to reduce the chance of failure and avoid unexpected energy accumulation during mining and tunneling operations. To further improve failure prediction, [45]found that strain localization and gradual deformation occur before large-scale collapse in brittle materials. Their laboratory-scale tests support the necessity of constant monitoring in geomechanical designs.

Recently, [46]showed that long short-term memory (LSTM) neural networks are better at predicting slope stability than conventional models like random forests and support vector machines. These deep learning models provided more dependable tools for managing geohazards by enhancing temporal sensitivity and sequence-based forecasting.

In order to forecast the long-term behavior of salt rocks, Hou et al. 2024 and Zhao et al. 2020 performed uniaxial cyclic loading creep tests and put forth a novel constitutive model[1,6]. They looked into how three cyclic conditions—average stress, half-amplitude, and cyclic period—affect the way salt rocks creep. Furthermore[15] created a nonlinear creep damage constitutive model for argillaceous salt rock using uniaxial creep experiments, which consider the accelerated creep phase as a combined loading and creep process. Test results show that the model, which applies to mudstone, high-purity salt rock, and mudstone-salt mixtures, performs better than conventional models for rapid creep.

This offers crucial theoretical backing for interbedded energy storage in salt caverns to operate safely. Power law, the logarithmic model, Maxwell's spring-dashpot model, and other conventional or analytical models have provided important insights into how salt rock deforms under continuous stress[6,16]. Although these traditional models have improved our knowledge of salt rock creep, they frequently fail to accurately predict long-term behavior. Promising substitutes for forecasting the creep response of salt rock are provided by recent developments in artificial intelligence (AI) and machine learning. The intricacies of rock deformation over a long period can be captured by sophisticated deep neural network (DNN) models, such as Neural Basis Expansion Analysis for Time Series (N-BEATS), Temporal Convolution Networks (TCN), Recurrent Neural Networks (RNN), and Transformers (TF). Following a resampling operation, these models can produce more accurate predictions of future deformation trends than traditional methods by training on experimentally acquired recorded creep data of axial strain responses with non-uniform sampling obtained from CSIRO's Geomechanics and Geophysics Lab for various confining stages at room temperature conditions[18,19].

In order to improve on earlier studies, this study compares the effectiveness of sophisticated deep-learning algorithms with conventional analytical models in forecasting the creep reaction of salt rock. In particular, we concentrate on the Frome rock salt formation, which is Devonian in age and may be used for long-term energy storage. We evaluate these models' relative predictive accuracy for creep behavior by examining axial strain data. Statistical metrics like the $R^2$ score and root mean squared error (RMSE), mean average error (MAE), mean average percentage error (MAPE in %), and symmetric mean average percentage error (SMAPE in %) will also be used to compare the predictions of DNN models with those of traditional models to assess their dependability.

Ultimately, this research seeks to enhance the accuracy and reliability of geomechanical models, contributing to the safe and efficient design of salt rock storage systems.

**Material and methods**

**Study area & multi-stage triaxial test (MSTL)**

For this study, the core samples were taken from the onshore FRDDH001 vertical well at the top edge of an interpreted salt diapir on the Jugurra Terrace, near the Fitzroy crossing in the Canning Basin, north of Western Australia[17]. This well intersects over 600 meters of early Paleozoic rock salt, evaporites, and dolomitic breccia[18,19].

This study used multi-stage triaxial (MSTL) tests on core samples extracted from salt-rock facies at around 1100 meters, with varying confining pressures. Creep monitoring was done at each pressure stage under a constant axial load value, with the deviatoric stress set to 75% of the estimated yield stress. Radial strain was minimal and ignored due to its high sensitivity to temperature fluctuations and low signal-to-noise ratio. In contrast, though temperature fluctuations affected the axial strain, it showed a clear trend due to its high signal-to-noise ratio[18–20]. A typical four-stage triaxial test on a rock salt sample involves loading and unloading stages at four different confining pressures. The third stage matches the native in situ condition of the rock, usually at 25 MPa for clean salt facies. The first and second stages occur at lower pressures than those estimated for the recovery depth, while the fourth stage is at a higher pressure. These stages are performed at an axial strain rate of $10^{-6}$ $s^{-1}$ with pressure changes between stages at a rate of 0.5 MPa/min[18,19]. Fig. 1 shows axial creep strain and temperature fluctuations recorded at each stage of the MSTL test for salt rock. The creep rate trend changes after 0 to 10 days, depending on the applied stress and rock salt facies. The slope of the creep curve at the end of a test can serve as a proxy for estimating the steady-state creep rate. The exponential (tertiary) creep and failure stage was not reached in the four tests conducted on rock salt samples at room temperature and under deviatoric stress, approximately 75% of the short-term yield stress at the given confining pressure.

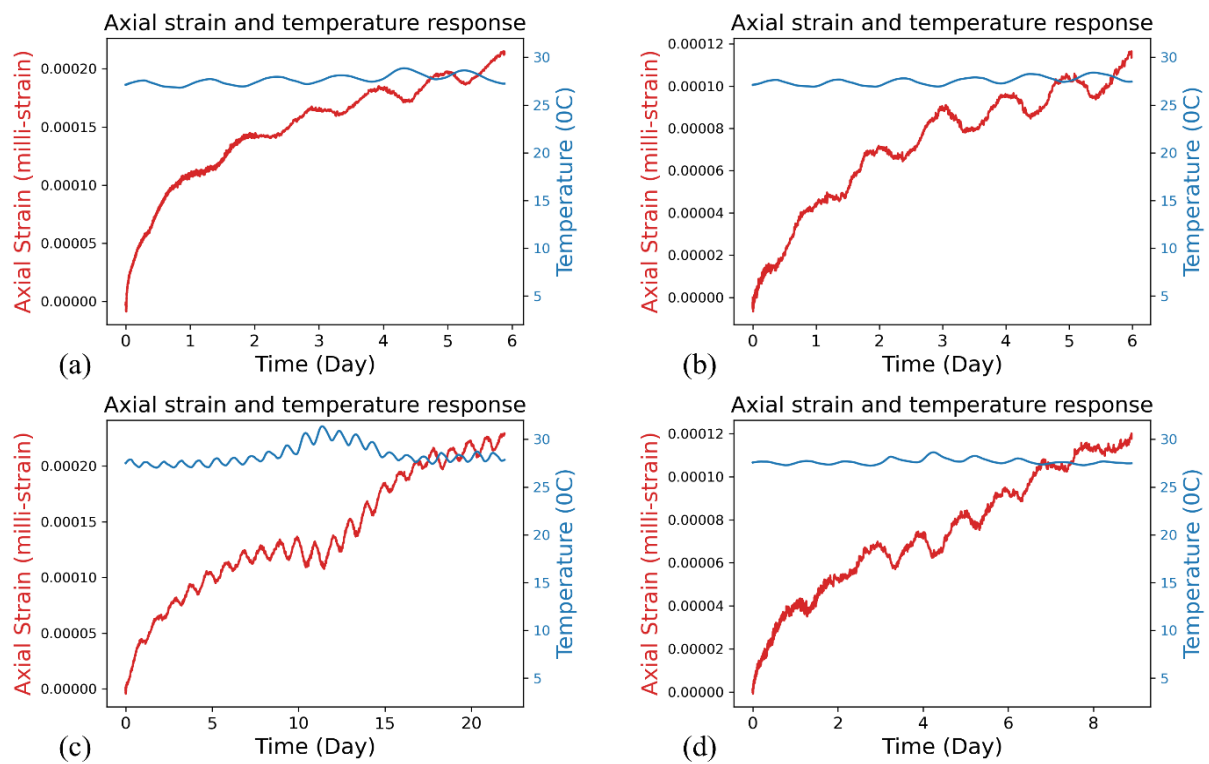

**Fig. 1:** The MSTL creep data of axial strain and temperature sensor responses over time, whereas axial strain (milli-strain) and temperature ($^0$C) with varying time (days) at various confining pressure levels such as for (a) 5MPa, (b) 15MPa, (c) 25MPa, and (d) 35MPa, respectively.

A detailed summary of the dataset, including its duration and frequency statistics, is provided in Table 1.

**Table 1**: Time series statistics including training and test sets.

| MSTL Stage | Confining stress levels (MPa) | Length duration (days) | Total length of the sample set (hour) | Training (hour) | Test (hour) |
|---|---|---|---|---|---|
| **Stage 1** | 5MPa | 5.87 | 141 | 100 | 41 |
| **Stage 2** | 15MPa | 5.95 | 143 | 100 | 43 |
| **Stage 3 (in-situ)** | 25MPa | 21.87 | 525 | 365 | 160 |
| **Stage 4** | 35MPa | 8.87 | 213 | 150 | 63 |

**Data collection and preparation**

High-quality salt rock facies samples, prepared with consistent dimensions, were used in this study. These samples were placed in an MSTL testing apparatus, where they were subjected to controlled axial loads and confining pressures in a room-level experimental setup. Fig. 2 shows how temperature variations (beyond diurnal) at about 300 hours affect (bump) the series 3 data set at 25 MPa. However, after testing statistical techniques like Granger causality and Augmented Dickey-Fuller (ADF), we first ignored temperature as a component in this study. Particularly at 25 MPa, there is no discernible correlation between temperature and axial strain. The only discernible pattern was that axial strain appeared to decrease as temperatures grew. We implemented a univariate approach, with the main elements of this study being axial strain and time (index). Future multivariate time series analysis could explore the possibility of adding temperature as an extra variable to enhance forecasting and pattern recognition.

The raw axial strain data from MSTL tests underwent meticulous preparation to ensure data integrity. Inconsistencies and outliers were removed, and the datasets featured non-uniform time sampling (in milliseconds) at various constant confining stages[4,8,9,17]. A resampling operation was performed to streamline the data, converting high-frequency data into uniform hourly intervals. This reduced data volume and computational load while preserving significant trends and patterns. Normalization techniques were also applied to standardize the datasets, ensuring consistency across all data points. The performance of predictive models was rigorously evaluated using validation metrics, allowing the identification of best-fitting parameters to ensure accurate predictions.

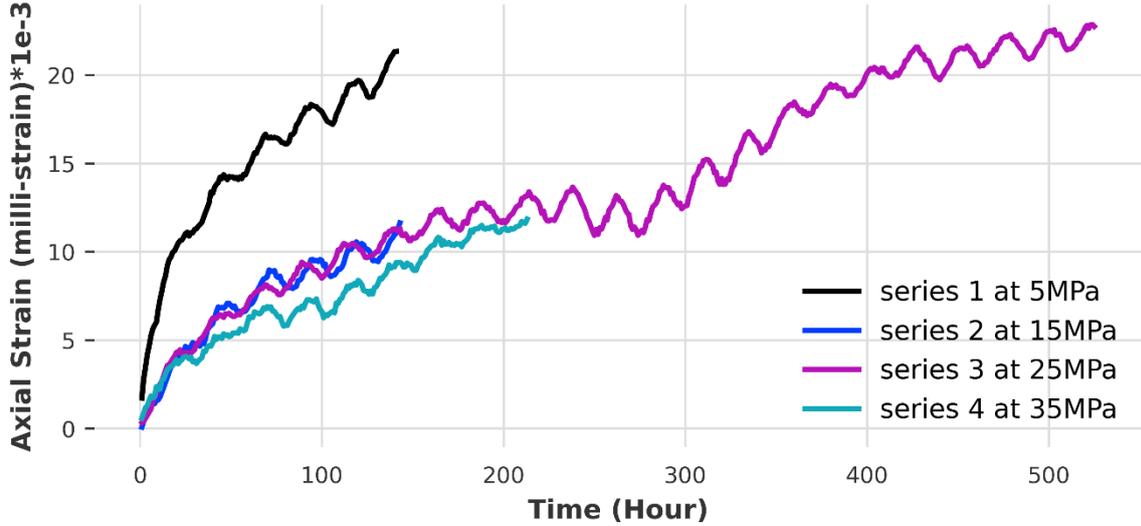

**Fig. 2:** Four-time series creep datasets for axial strain responses over time with different time stamp lengths (hours) at various pressure levels.

**Predictive modeling architecture**

In this study, we applied DNN and statistical baseline models to forecast the time-dependent deformation of salt rock, focusing on the axial strain response under varying pressures and temperatures. The resampled time series data, taken at fixed one-hour intervals, incorporated four time series datasets of axial strain responses over time. Before applying DNN and traditional forecasting models, we performed Seasonal-Trend decomposition (STL) using LOESS to separate the data into its original, trend, seasonal, and residual components[21–23]. Key statistical tests, such as Granger causality, ADF, and wavelet coherence plots (WCF), were applied[24–26]. The Granger test analyzed the influence of temperature on axial strain over time, while the ADF test examined the data's stationarity.

We utilized various DNN models for forecasting, including N-BEATS, Temporal Convolutional Networks, Transformers, and Recurrent Neural Networks. Statistical models such as Exponential Smoothing (ES), Trigonometric Box-Cox, ARMA (autoregressive moving average), trend and seasonal components (TBATS), automatic autoregression integrated moving average (AutoARIMA), and Theta were employed[27–31]. Analytical models—power law, Kelvin, Burger, logarithmic, and Maxwell spring dashpot—were used to complement the deep learning forecasts for predicting axial strain over extended periods. Model performance was evaluated using key metrics such as RMSE, MAE, MAPE (%), and SMAPE (%)[1,32]. The results were further validated using Multiple Comparisons with the Best (MCB) tests based on mean rank comparisons[32].

This comprehensive workflow, involving STL, causality, statistical tests, and rigorous model evaluation, ensured robust and reliable forecasting. Our findings provide valuable contributions to geoscience applications with statistically validated and effective models for real-world forecasting. Our proposed workflow summarizing these steps is presented in Fig. 3.

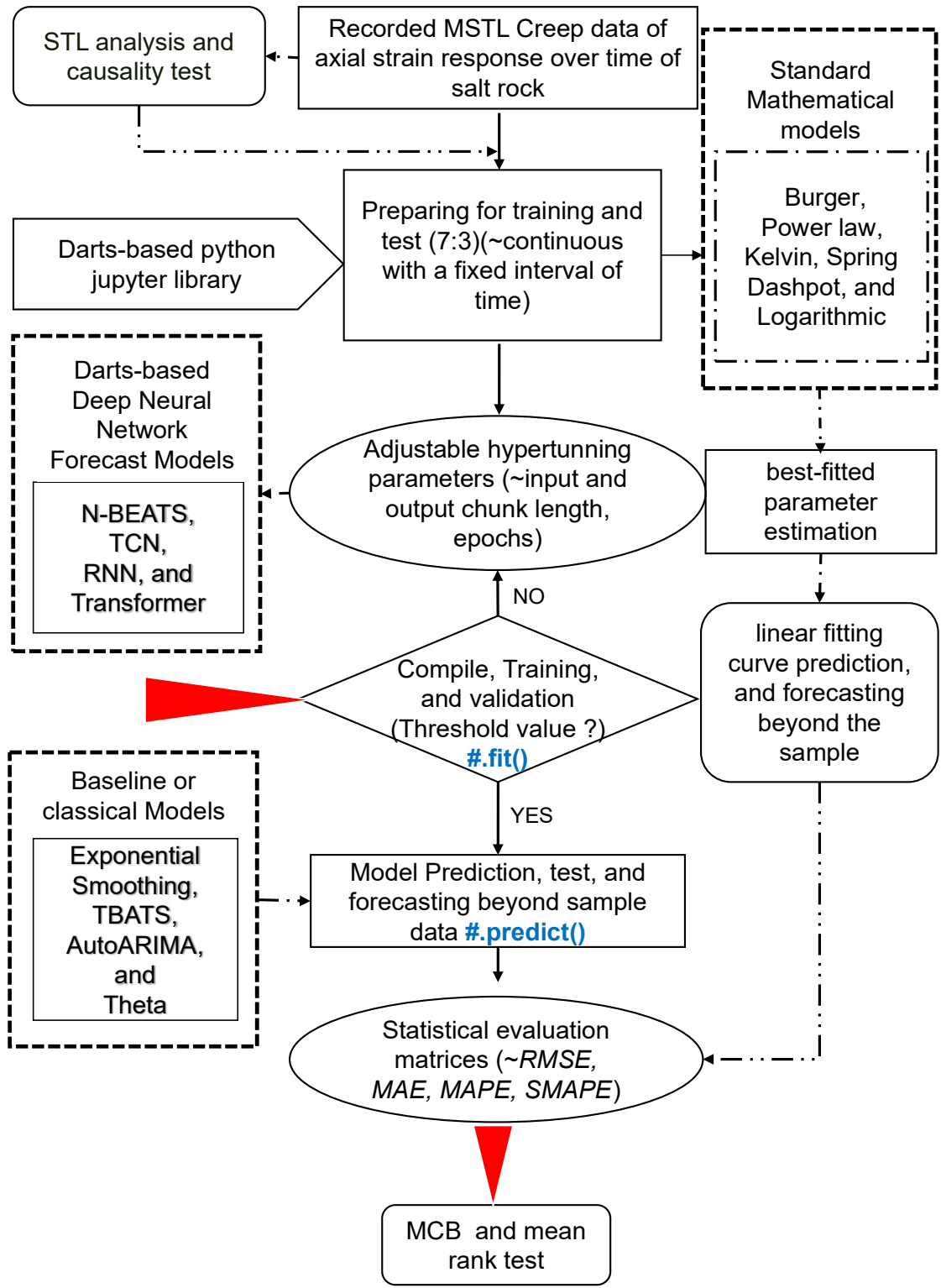

**Fig. 3:** Proposed workflow for predictive modeling of salt rock's creep deformation.

**Time series decomposition & statistical tests**

Observing and assessing seasonality, trends, and noise is possible when examining a time series's temporal data[33–36]. Each element can be investigated separately to better comprehend the data. However, two more elements can be seen when examining the data as a whole: stationarity and autocorrelation. It is essential to comprehend these elements while choosing a time series model to get precise forecast outcomes. Seasonal-trend decomposition with LOESS is a reliable technique for breaking down time series, where LOESS is a nonlinear relation estimation approach. The STL method estimates the trend components—a flexible estimate of the underlying pattern in the data—using a locally weighted regression technique called LOESS. The time series data Y(t) at a specific time (t) is broken down by STL into three components: a trend component (T(t)), a seasonal component (S(t)), and a remainder or residual component (R(t)). The findings are displayed as follows in Eqn. (1)[32,36,37]:

$$Y(t) = T(t) + S(t) + R(t), \quad t = 1,2,3 \ldots N \qquad (1)$$

Where Y (t) is the actual time series dataset with the function of varying time *t = 1,2,3…..N.*

ADF test, the Granger causality test, and WCP to analyze time series datasets comprehensively[24,38]. The ADF test ensures stationarity in the analyzed time series, which is crucial for reliable time series modeling. The Granger causality test identifies causal relationships between series, enhancing forecasting accuracy. Wavelet coherence plots visualize correlations across time and frequency domains, providing deeper insights into dynamic interactions. These methods offer a robust way to understand dependencies, improve model performance, and inform decision-making in various applications such as economics, finance, environmental, and geoscience studies, especially related to long-term storage of energy—wavelet coherence is particularly used to detect common time-localized oscillations in non-stationary signals. If one time series influences another, one may use the phase of the wavelet cross-spectrum to identify the relative lag within the time series datasets.

**Time series forecasting (TSF)**

A time series consists of chronologically ordered observations recorded at fixed intervals, with the forecasting problem focusing on predicting future values based on past observations (lags). Given historical data X = $\{x_1, x_2, x_3, \ldots.. x_T\}$, and a desired forecasting horizon H, the goal is to predict subsequent values $\{x_{T+1}, x_{T+2}, x_{T+3}, \ldots \ldots \ldots x_{T+H}\}$.

Being $\hat{X}$ $\{x_1, x_2, x_3, \ldots., x_T\}$ is the vector of predicted value, While minimizing prediction error (PE) is defined as [27,28,31]:

$$PE = \sum_{i=1}^{h=H} |x_{T+i} - \hat{x}_{T+i}| \tag{2}$$

Time series can be classified as univariate (single observation) or multivariate (multiple observations). This study focuses on univariate time series analysis of axial strain recorded over time.

In recent decades, machine and deep learning techniques have gained prominence for time series forecasting (TSF), serving as advanced alternatives to traditional statistical methods. TSF is crucial in various fields, including energy consumption, financial forecasting, and anomaly detection[28]. Effective TSF models must address trends, seasonal variations, and correlations among closely observed values. Deep neural networks (DNNs) have emerged as powerful tools for TSF, capable of capturing complex non-linear relationships that traditional models cannot. This study evaluates four DNN models—N-BEATS, TCN, RNN, and Transformer—alongside four statistical baseline models—ES, TBATS, Auto ARIMA, and Theta—specifically for forecasting axial strain deformation in geoscience applications.

We aim to provide a reliable benchmark for future studies by comparing these models without requiring refined architectures tailored to specific problems. The proposed domain-independent approaches are intended to guide researchers in effectively tackling forecasting challenges in various contexts.

**Global forecasting DNN models**

The DNN models—N-BEATS, TCN, RNN, and Transformers offer unique advantages and challenges for modeling nonlinear time series. N-BEATS excels in decomposing time series data into basis functions, and TCNs leverage dilated causal convolutions for capturing temporal dependencies. RNNs maintain a hidden state for sequence modeling, and Transformers utilize self-attention mechanisms for parallel processing and capturing long-range dependencies. In this study, we considered the following deep learning methods for forecasting salt rock deformation description given below;

**N-BEATS**

N-BEATS[25] is a deep neural network architecture designed to forecast time series data by decomposing it into basis functions. This approach enables the model to effectively learn and combine complex patterns for accurate predictions. The N-BEATS model is structured with multiple layers that handle trend and seasonality components, enhancing its performance across various time series datasets. Its interpretability allows time series modeling to generate better forecasts.

**TCN**

TCN[22,39,40] utilizes dilated causal convolutions to efficiently capture temporal dependencies in sequences. Unlike RNNs, TCN relies on these convolutions to consider only past information, thus avoiding gradient explosion and vanishing issues. Due to their parallel processing capabilities and stable training dynamics, TCNs are particularly effective for modeling long-term dependencies and handling large datasets.

**RNN**

RNNs[24,32] model is designed for sequence modeling, maintaining a hidden state that captures temporal dependencies across inputs. This characteristic allows RNNs to process sequences of varying lengths and is crucial for tasks like time series prediction. However, when dealing with long sequences, RNNs face challenges such as gradient vanishing and explosion.

**TF**

The Transformer model[42] is renowned for its self-attention mechanism, which allows it to weigh distinct parts of an input sequence differently. This feature is beneficial for capturing long-range dependencies and handling complex temporal patterns in time series data. Although Transformers were initially developed for natural language processing, their ability to process sequences in parallel and capture global dependencies makes them valuable for time series forecasting. Various adaptations have been made to tailor Transformers for sequential data, improving their performance in this domain.

**Statistical baseline models**

**ES**

ES is a basic statistical technique used to smooth out time series data that often exhibits long-term and short-term variability. It applies uniform or exponential weights to past observations, making long-term variability more evident as short-term variability (noise) is removed. Unlike the simple moving average, this model assigns weights that decrease exponentially for older data points. This method is widely used for forecasting because it effectively captures short-term fluctuations and trends while smoothing noise and irregularities[23,26].

**TBATS**

TBATS[23-24,26] is a time series forecasting model that can handle intricate seasonal patterns, especially ones with several seasonalities. The data is particularly helpful when non-linear trends, high-frequency seasonality, and variable degrees of seasonality are present throughout the series. ARMA is used to model residuals, while Box-Cox transformations stabilize the variance. It incorporates multiple components such as trend, seasonality, and irregular patterns, whether yearly, weekly, daily, or hourly. It offers robust performance and incorporates automatic model selection and parameter estimation.

**Auto ARIMA**

AutoARIMA is a model class that predicts a time series's future values based on its past lagged values (lags) and lagged errors. It is effective for non-seasonal time series exhibiting patterns, not random white noise. ARIMA is agile in adapting to various types of time series data and can be trained on smaller datasets. However, ARIMA may not handle data with large mean shifts and nonlinear time series. AutoARIMA is an automated approach to fitting ARIMA models to time series data. It automatically selects optimal parameters (p, d, q) based on statistical metrics like AIC (Akaike Information Criterion) and BIC (Bayesian Information Criterion), which are statistical measures used to evaluate and select models, simplifying the forecasting process. The model with the lowest AIC and BIC values is considered the best[21,26,29].

**Theta**

In time series analysis, the Theta model[29] is a forecasting technique that uses ES[26,27]. For future trend prediction, it combines level and trend data components. A straightforward linear model and a decomposition technique are used to create this potent time series forecasting technique. The Theta model begins by breaking down a given time series Y(t) into its component parts, such as the trend component, T(t), seasonal component, S(t), and residual component, R(t), as was previously described in Eqn. (1). Using the Theta (θ) coefficient to alter the trend component is the main concept of the Theta model during the Theta Transformation. The Theta transformation modifies the trend to change the series' curvature:

$$Y_\theta(t) = \theta \times T(t) + (1 - \theta) \times Y(t) \tag{3}$$

By strengthening or weakening the trend, this equation alters the original time series. The series is detrended and deseasonalized when θ=0, emphasizing the error and seasonal components. In this instance, it generates a series in which the trend is eliminated, leaving only the noise and seasonal components, like

$$Y_0(t) = Y(t) - T(t) \tag{4}$$

When θ=2, the trend is overstated, which may make it easier to identify the long-term trend. In this instance, it results in a series that highlights a tendency, like

$$Y_2(t) = 2.T(t) - S(t) - R(t) \tag{5}$$

A basic forecasting technique (often ES or linear extrapolation) is used for every θ-transformed series. These two equations (4) and (5), where θ=0 and 2, are combined to get the final forecast (Ƴ(t+h), which is usually derived by averaging the forecasts from several θ-transformed series:

$$\hat{Y}(t+h) = \frac{1}{2}[\hat{Y}_0(t+h) + \hat{Y}_2(t+h)] \qquad (6)$$

Where h is the forecasting horizon, and $\hat{Y}_\theta(t+h)$ is the forecasted value for time *t+h*.

With this method, the advantages of both the accentuated trend line (θ=2) and the deseasonalized trend line (θ=0) are combined. It is a statistical baseline model that works well for time series forecasting because it strikes a fair mix between capturing the long-term trend and short-term seasonality. It is crucial to comprehend that the model generates the final forecast by transforming the trend and seasonal components in various ways. The Theta model is therefore a straightforward yet effective time series forecasting method[29].

A crucial tool for producing accurate forecasts based on past data trends is offered by the four statistical baseline forecasting models mentioned above.

**Analytical models**

The viscoplastic behavior of materials under creep circumstances can be understood and predicted with the help of analytical models[4,8,14,35-36]. Several researchers can learn more about the properties of material deformation by fitting these models to experimental data, which improves the precision of long-term trajectory predictions. As earlier discussed, the creeping phenomenon as a function of time, stress, temperature, and material qualities is expressed using empirical or phenomenological equations, which are commonly used to predict creep responses over time. Because it captures the nonlinear relationship between stress and strain rate, the power law model is the most commonly used for understanding the nature of the viscoelastic response of creep. The creep deformation over time under a given stress is forecast by the various analytical models, such as the power law, extended Kelvin, Burger, or spring dashpot model.

In order to effectively depict viscoplastic effects, Burger's model combines aspects of the Maxwell and Kelvin-Voigt models. These analytical models are appropriate for complex viscoelastic behavior since they take into consideration both instantaneous and time-dependent viscous reactions[2,4,8]. Time-dependent elastic and viscous trends are also captured by the Kelvin-Voigt model; these trends are especially pertinent for materials with delayed elastic responses, such as polymers and biological tissues. Furthermore, by simulating the time-dependent deformation under various confining pressure circumstances, the Maxwell model—which excels at modeling viscoplastic behavior—is distinguished by its spring-dashpot configuration[1,2,4,10,35-38]. The stimulation of creep processes is often facilitated by the combination of the logarithmic and power law models. The axial strain response over time, ϵ(t), as a function of time t, can be expressed via Kelvin, Burgers, Maxwell, Power Law, and Logarithmic analytical models. Table 2 summarizes how these models use empirical creep constants for long-term predictions. Despite being a simplified depiction of real-world creep, the power law model might not adequately account for the intricacies of various materials and circumstances. For salt rock deformation, the related creep data on axial strain responses, as shown in Table 2, were previously utilized to fit popular creep models[7,16,38].

**Table 2:** A list of analytical models for creep responses for long-term forecasting.

| Model name | Description | Equations |
|---|---|---|
| Kelvin | The Kelvin model, with a parallel spring and dashpot in series with an additional spring. | $\varepsilon(t) = A_1 + A_2\, e^{-k_1 t}$ where $A_1$ represents the initial strain contribution from the additional spring, $A_2$ represents the strain contribution that decays exponentially over time, and $K_1$ is a rate constant related to the dashpot and spring in the Kelvin element. |
| Power law | The Power Law model describes the strain response with a power law relationship. A power-law model assumes a viscous flow-like deformation with time. | $\varepsilon(t) = C\, t^k$ where C is a material constant, and k, sometimes written as n, is the power law exponent that defines the nonlinearity |
| Burger | Burger's model is equivalent to the (non-elastic creep component) Maxwell spring-dashpot models in series. | $\varepsilon(t) = B_1\,(1 - e^{-k_2 t}) + B_2\, t$ where $B_1$ represents the strain contribution from the Kelvin element, $K_2$ is a rate constant related to the Kelvin element, and $B_2$ represents the linear strain contribution from the Maxwell element. |
| Logarithmic | The Logarithmic model describes the strain response with a logarithmic relationship. | $\varepsilon(t) = D_1 + D_2\, t + D_3 \log(1 + k_3\, t)$ where $D_1$ represents the initial strain, $D_2$ represents the linear strain contribution over time, $D_3$ represents the amplitude of the logarithmic term, and $K_3$ is a rate constant related to the logarithmic term. |
| Spring- dashpot/Maxwell | In viscoelasticity, a combination of springs and dashpots represents the material elements of the Maxwell and Kelvin-Voigt models, representing both elastic and viscous components. The spring-dashpot model described four significant parameters to describe the material | $\varepsilon(t) = \varepsilon_0 + \sigma_0 \left[\frac{1}{E} + \frac{1}{\eta}\exp\left(-\frac{1}{\eta}t\right)\right]$ where $\varepsilon(t)$ is the total axial strain at varying time (t), $\varepsilon_0$ is the initial strain, $\sigma_0$ is the applied stress, E is Young's modulus (elastic modulus), and $\eta$ is the viscosity of the material. Over a long period, the viscous component decays to zero, leaving only the initial and elastic strain components. |

| | response, i.e., time (t), epsilon (t=0), Young modulus (E), and viscosity. | |

**Model implementation and training process**

We used deep learning models to predict axial strain time series data methodically (See Darts library developed with the Python programming language). To enable accurate performance evaluation, data preparation entailed splitting the dataset into training and testing sets (Fig. 4). The Number of Epochs, input, and output chunk lengths were two important hyperparameters that were tuned for the model's performance. To enhance DNN convergence, we scaled each training time series to a range of 0 to 1 once the data was separated. Using a Multi-Input Multi-Output (MIMO) approach, which has been demonstrated to perform better than single-output techniques by preventing error accumulation and increasing computing efficiency, training examples were produced[24,27]. For example, Fig. 5 shows input-output pairings with a forecast horizon of 12 and a history length of 24 generated by a fixed-size sliding window. To determine the ideal training setup, we also experimented with different batch sizes, learning rates, and normalization strategies; the results are shown in Table 3.

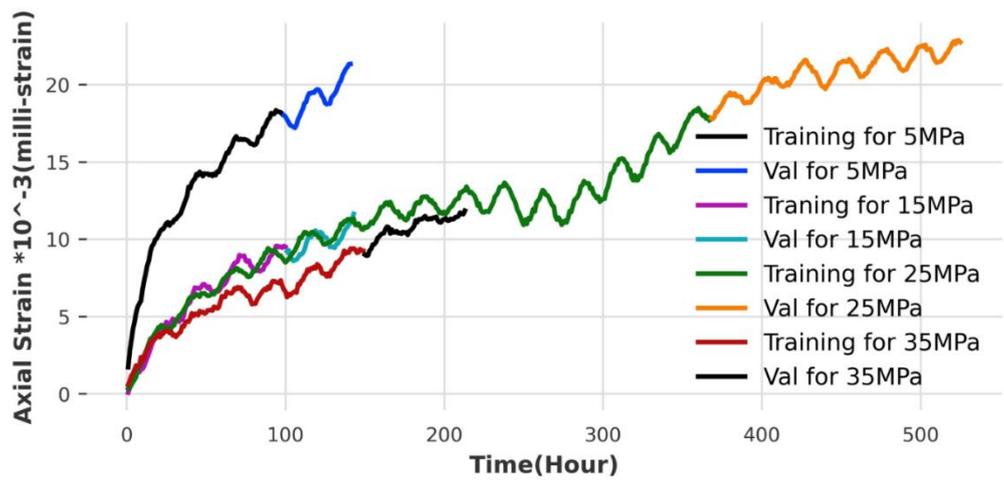

**Fig. 4:** Split data for training (70%) and validation test (30%) for axial strain response time series analysis at different pressure levels.

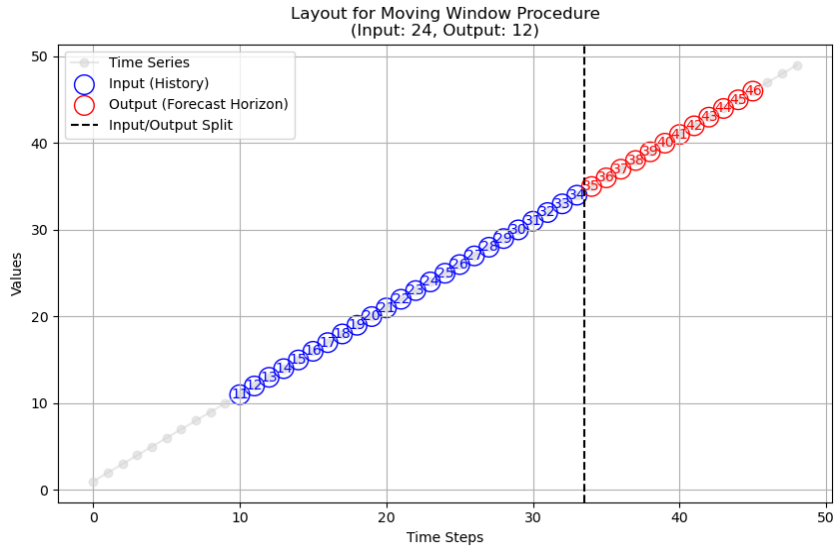

**Fig. 5:** A moving window technique example that gathers input/output examples for model training. The models are given an instance in this example with a forecasting horizon of 12 values and an input history window of length 24.

For evaluating model performance and directing modifications, the test dataset was essential. The outcomes showed how deep learning models could capture intricate temporal patterns. In order to forecast axial strain responses, we used DNN models (N-BEATS, TCN, RNN, and Transformers), statistical baseline models (ES, TBATS, Auto ARIMA, and Theta), and analytical models (power law, Burger's, Kelvin's, logarithmic, and spring dashpot models).

To improve performance, grid search and cross-validation were used to adjust the hyperparameters of each model, which was customized for the time series data.

**Table 3:** Darts-based default hypertunning parameters for DNN models used in this study.

| SN | Model Name | default hypertunning parameters in a Darts-based environment | |
|---|---|---|---|
|  |  | Input variable | Input action |
| 1 | N-BEATS | Number of stacks | 30 |
|  |  | Generic architecture | true |
|  |  | Number of layers | 4 |
|  |  | Number of blocks | 1 |
|  |  | Layer width | 256 |
|  |  | Expansion coefficients dim | 5 |
|  |  | Trend polynomial degree | 2 |
|  |  | dropout | 0.0 |
|  |  | Activation function | Relu |

|   |   | Output chunk shift | 0 |
|---|---|---|---|
| 2. | TCN | Output chunk shift | 0 |
|   |   | Kernal size | 3 |
|   |   | No of layers | none |
|   |   | dilation base | 2 |
|   |   | Weight norm | false |
|   |   | dropout | 0.2 |
| 3. | RNN | Output chunk shift | 0 |
|   |   | model | "RNN" or "LSTM" or "GRU" |
|   |   | hidden dim | 25 |
|   |   | RNN layer | 1 |
|   |   | dropout | 0.0 |
|   |   | hidden fc sizes | none |
|   |   | Batch size | [32, 64] |
| 4. | Transformer | Output chunk shift | 0 |
|   |   | Kernal size | 3 |
|   |   | No of filter | 3 |
|   |   | No of layer | none |
|   |   | Dilation base | 2 |
|   |   | Weight norm | False |
|   |   | Dropout | 0.2 |
|   |   | Random state | 0 |

**Model performance**

Four statistical criteria were used to assess each model's performance[4,14,21,22];

$$RMSE = \sqrt{\frac{1}{n}\sum_{i=1}^{n}(y_i - \hat{y}_i)^2} \tag{7}$$

$$MAE = \frac{1}{n}\sum_{i=1}^{n}|y_i - \hat{y}_i|, \tag{8}$$

$$MAPE\ (in\ \%) = \frac{1}{n}\sum_{i=1}^{n}\left|\frac{y_i - \hat{y}_i}{y_i}\right| \times 100, \tag{9}$$

$$SMAPE\ (in\ \%) = \frac{1}{n}\sum_{i=1}^{n}\left|\frac{y_i - \hat{y}_i}{(|y_i| + |\hat{y}_i|)/2}\right| \times 100 \tag{10}$$

Where $y_i$ represents the actual value of axial strain and $\hat{y}_i$ represents the forecast value of axial strain obtained from different forecast models, and n represents the length of the series. By definition, the lower the values of these metrics, the better the forecasting models.

The output of the evaluation metrics is summarized in Tables 5 and 6 later in the result section, which detail model performance across four different confining pressures. It is important to note that discrepancies between model outputs and actual values arise from two primary sources: sample error, which can be mitigated through effective outlier processing, and model error, which is linked to the model's generalization ability, which can limit predictive accuracy. Forecast accuracy is a critical measure that reflects how well the model fits the actual data, underscoring the importance of refining models to improve their predictive capabilities.

**Results**

**Preliminary data analysis**

The STL decomposition of the axial strain dataset (Figs. 6a and b) exhibited minimal seasonal variation, with significant shifts in trend and residual components after the time data were converted to hourly intervals. At 5MPa confining pressure, the residual component displayed pronounced fluctuations, suggesting an influence of pressure on the strain response.

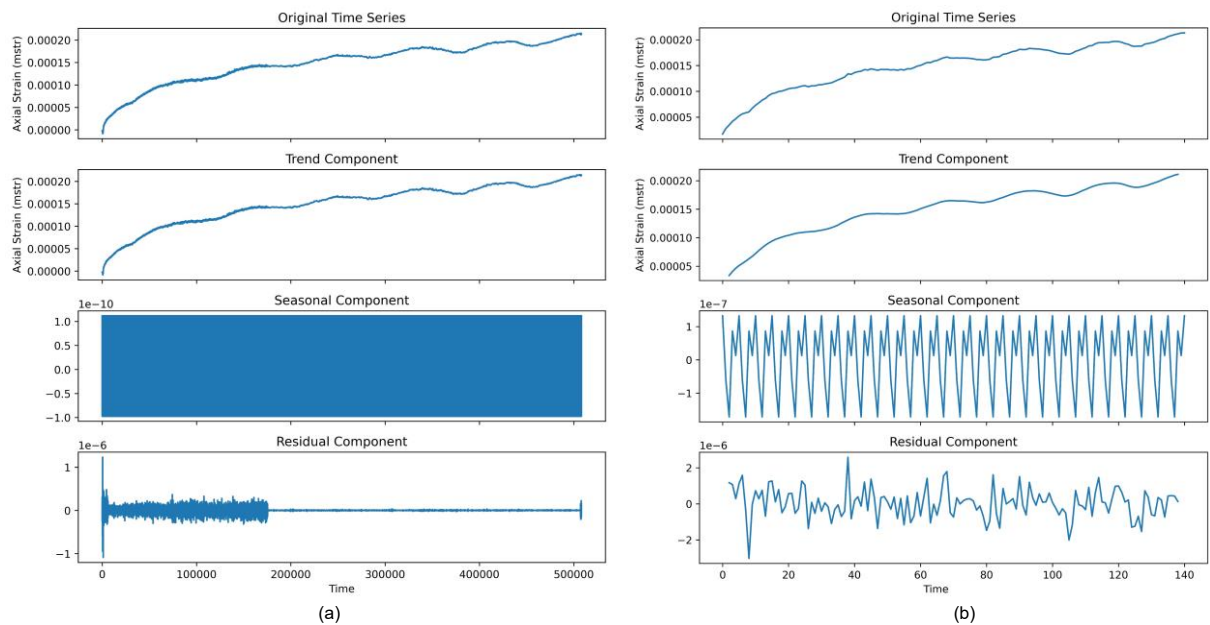

**Fig. 6:** An example of STL decomposition that decomposed data into original, trend, seasonal, and residual components for (a) frequency in seconds and (b) frequency in hours, respectively, at a pressure level of 5MPa.

Granger causality analysis (Figs. 7a and b) revealed no significant causal relationship between axial strain and temperature across most conditions. However, at 5 MPa, a p-value of 0.016574 confirmed a significant causal link between the variables. The ADF test (Fig. 7c) further validated the stationarity of the datasets, with all p-values falling below the 0.05 threshold, indicating the data's suitability for reliable predictive modeling (Table 4). WCP (Fig. 7d) identified clear cyclic patterns and external correlations within the dataset, particularly in lower frequency bands. These coherence magnitudes, ranging between 0 and 1, pointed to strong correlations in specific frequency ranges, demonstrating external influences on axial strain behavior over time. Such findings highlight the dataset's potential for accurate forecasting in future applications.

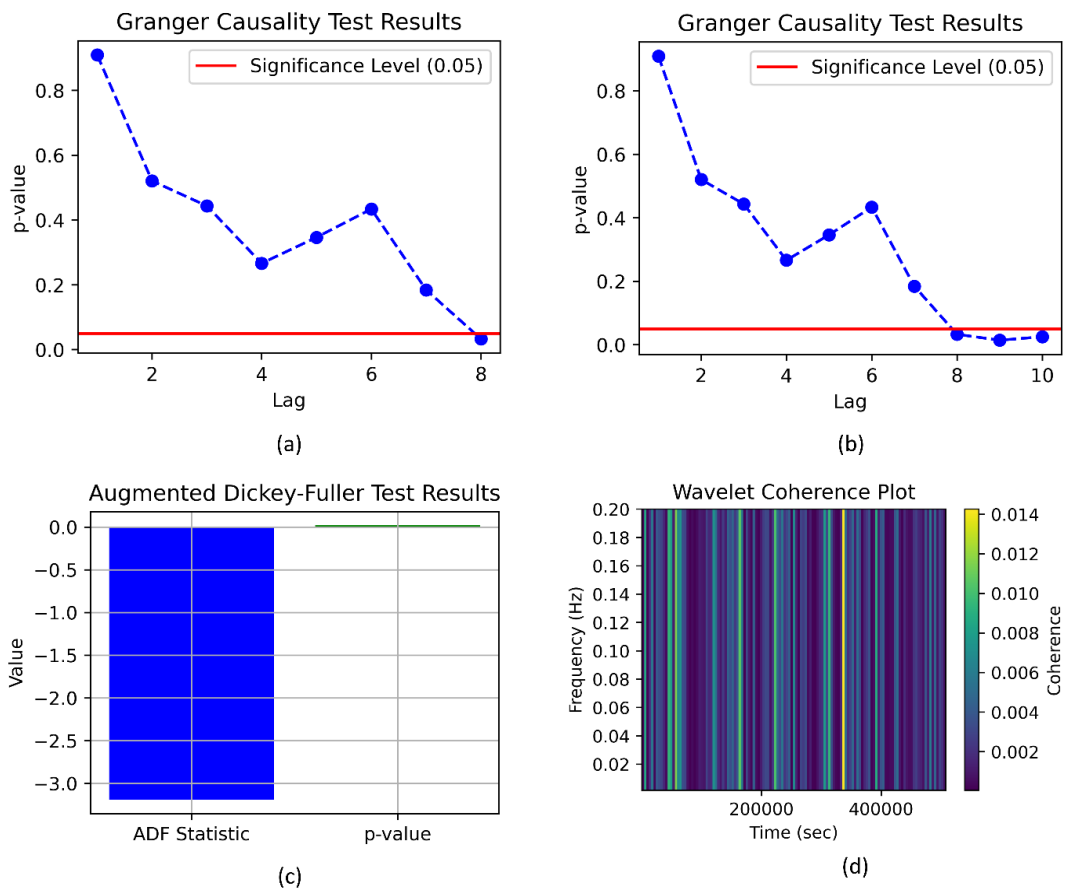

**Fig. 7:** An example of a time series for axial strain data is a causality test with different statistical tools such as (a) granger test, (b) granger test with increases lags more data points below or lies at significant levels (p-value less than 0.05), (c) ADF test for stationarily or non-stationary, and (d) WCP respectively at confining pressure condition of 5MPa.

**Table 4:** Statistical analysis summary with STL and causality test for creep data of axial strain responses with varying time at various confining pressure level conditions.

| Creep datasets | Seasonal/non-seasonal | Causality analysis | p-value (reference value <0.05) | Stationary/non-stationary |
|---|---|---|---|---|
| 5MPa | hardly observed any seasonality in the data | There is hardly any causality between strain and temperature data | 0.016574 | stationary |
| 15MPa | same | same | 0.037656 | stationary |
| 25MPa (in-situ) | same | same | 0.021693 | stationary |
| 35MPa | same | same | 0.017894 | stationary |

**Outcome from DNN models**

In our analysis of DNN models for time series forecasting, N-BEATS emerged as the top performer, achieving the lowest RMSE, MAE, MAPE (%), and SMAPE (%) values across all evaluated metrics. This model effectively captured complex temporal dependencies, making it suitable for short to medium-term forecasts. Specifically, the average performance metrics for N-BEATS varied as follows: RMSE (0.33–0.95), MAE (0.28–0.96), MAPE (1.45 %–4.54 %), and SMAPE (1.46%–4.62%) at various confining pressure levels (5 -35 MPa) as summarized in Table 5. In contrast, the TCN ranked second, with slightly higher average error metrics (RMSE: 0.54–1.35, MAE: 0.47–1.17, MAPE: 2.85–6.28, SMAPE: 2.88–6.03). While it outperformed other models in capturing longer-term trends, the Transformer model showed moderate performance, hindered by the limited dataset size. Although RNNs like LSTM and GRU outperformed baseline models, they did not match the accuracy of N-BEATS and TCN.

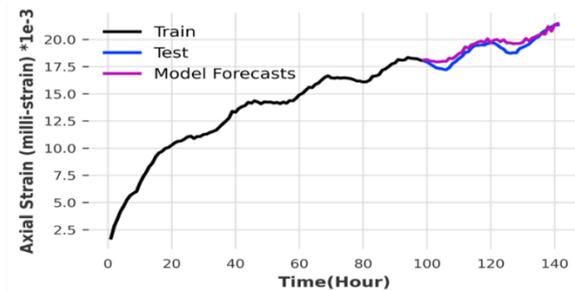

(a)

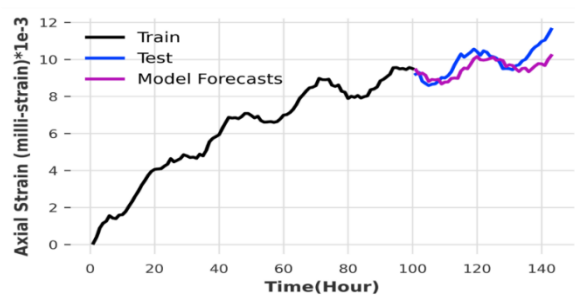

(b)

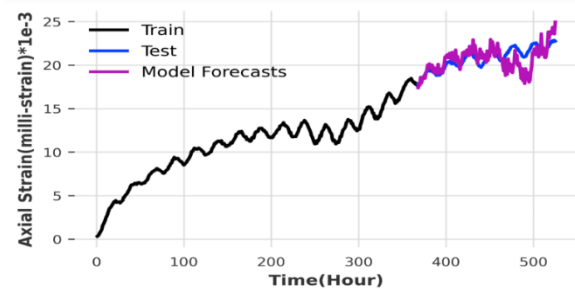

(c)

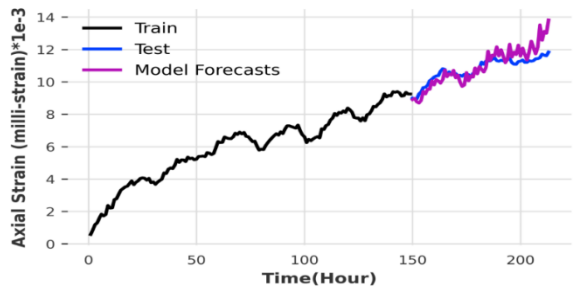

(d)

**Fig. 8:** The proposed N-BEATS model forecasts for (a) 5MPa, (b) 15MPa, (c) 25MPa, and (d) 35MPa, respectively, MSTL creep data of axial strain data, whereas the axial strain is in milli-strain and the time stamp is in hours. The best prediction performance was achieved using evaluation metrics such as RMSE, MAE, MAPE (%), and SMAPE (%).

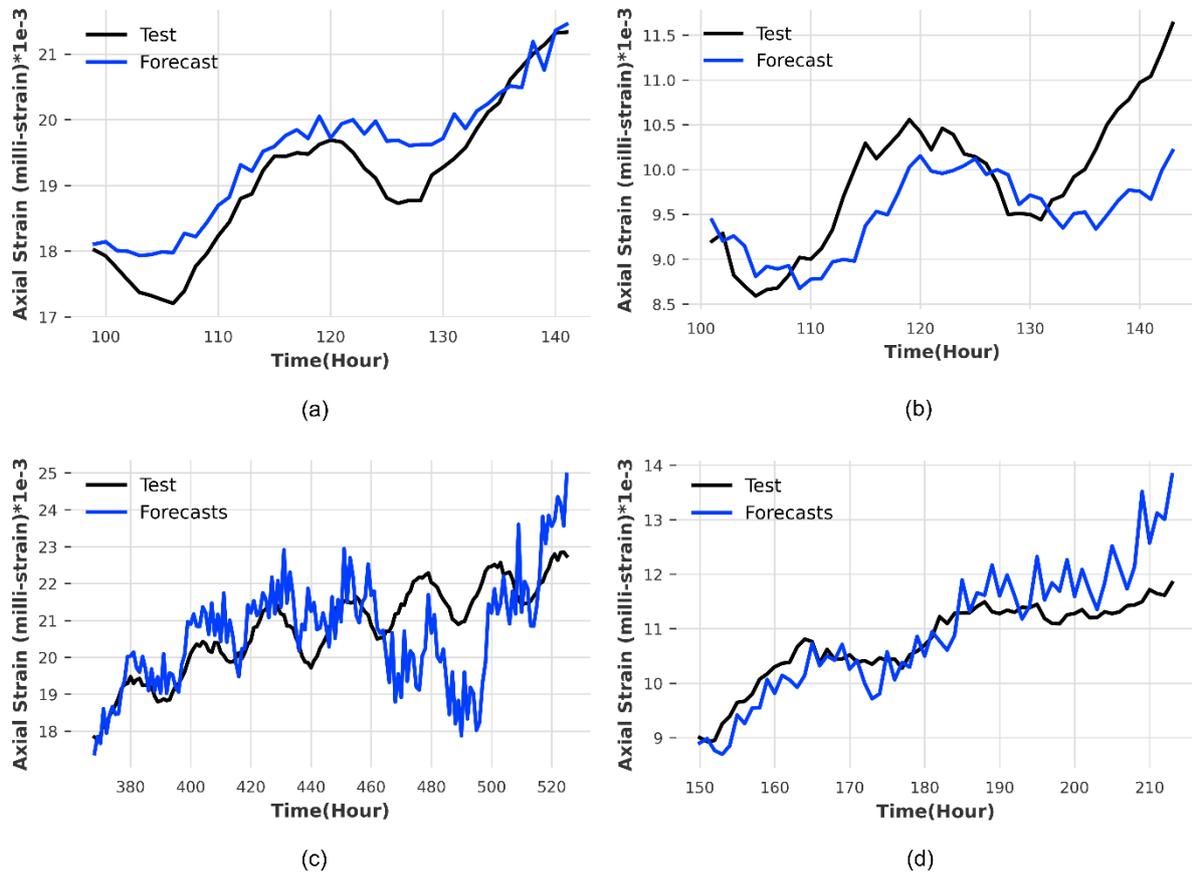

(a) (b)

(c) (d)

**Fig. 9:** The proposed N-BEATS model validation between forecasts against test data for (a) 5MPa, (b) 15MPa, (c) 25MPa, and (d) 35MPa, respectively, MSTL creep data of axial strain data, whereas the axial strain is in milli-strain and the time stamp is in hours. The best prediction performance was achieved using evaluation metrics such as RMSE, MAE, MAPE (%), and SMAPE (%).

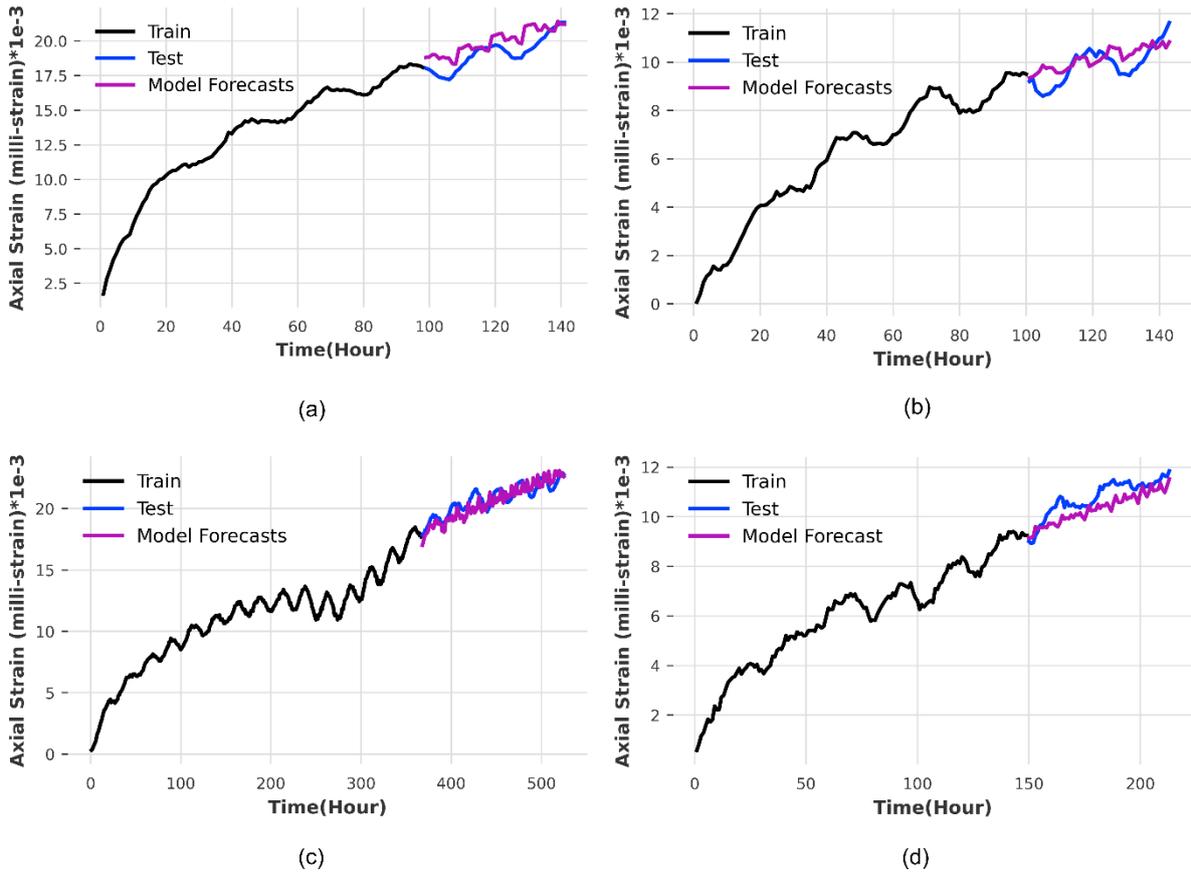

**Fig. 10:** The TCN model forecasts for (a) 5MPa, (b) 15MPa, (c) 25MPa, and (d) 35MPa, respectively, MSTL creep data of axial strain, whereas the axial strain is in milli-strain and the time stamp is in hours. The best prediction performance was achieved using evaluation metrics such as RMSE, MAE, MAPE (%), and SMAPE (%).

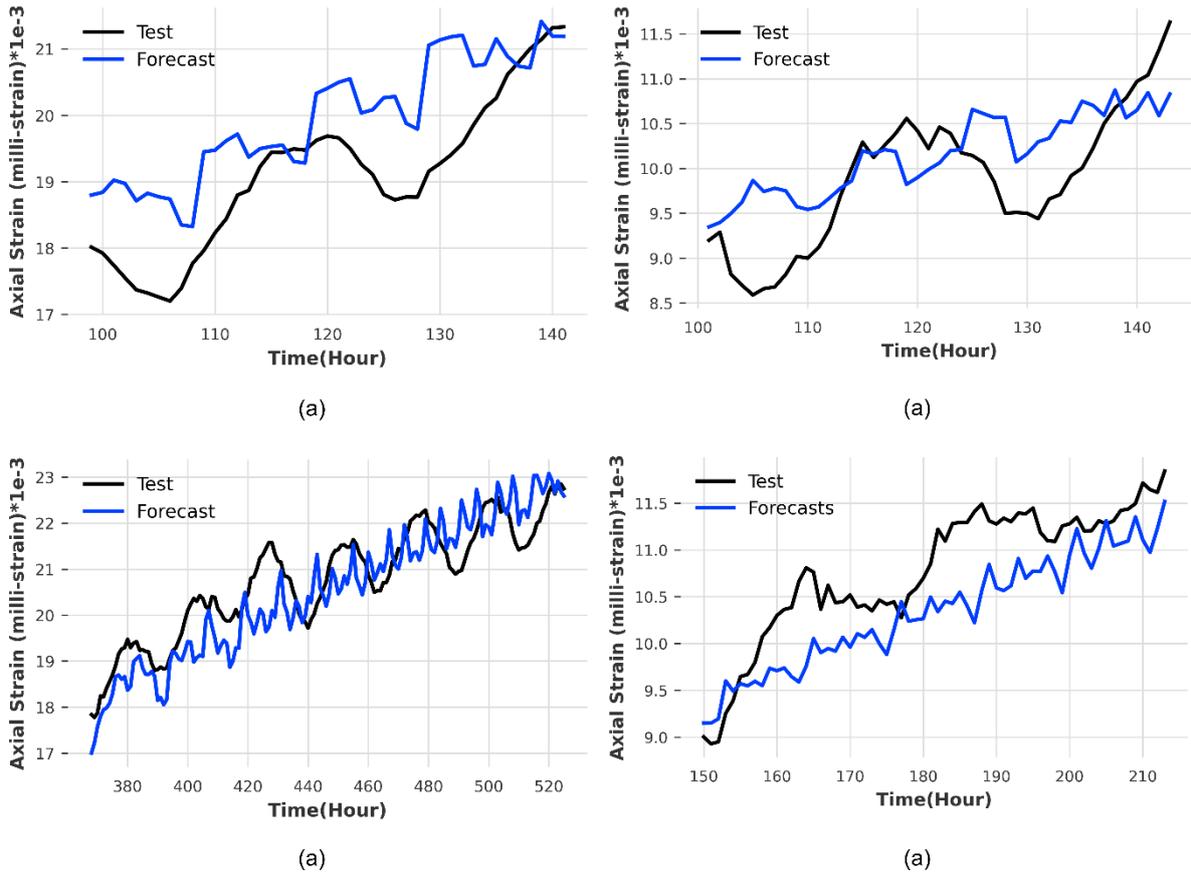

**Fig. 11:** The TCN model validation between forecasts against test data for (a) 5MPa, (b) 15MPa, (c) 25MPa, and (d) 35MPa, respectively, MSTL creep data of axial strain data, whereas the axial strain is in milli-strain and the time stamp is in hours. The best prediction performance was achieved using evaluation metrics such as RMSE, MAE, MAPE (%), and SMAPE (%).

**Table 5**: Summary of performance metrics and adjustable training parameters of deep learning models for forecasting the axial strain response creep datasets at various confining stress levels (best highlighted in bold)

| Stages | DNN Forecast Models | Epochs | Training loss | Input chunk length | Output chunk length | SMAPE (%) | MAPE (%) | RMSE | MAE |
|---|---|---|---|---|---|---|---|---|---|
| **5MPa** | **N-BEATS** | **100** | **0.0454** | **24** | **12** | **1.46** | **1.45** | **0.332** | **0.287** |
| | TCN | 400 | 2.260 | 24 | 12 | 6.03 | 6.28 | 1.352 | 1.177 |
| | RNN | 400 | 0.892 | 24 | 12 | 25.08 | 22.16 | 4.437 | 4.283 |
| | TF | 400 | 0.450 | 24 | 12 | 13.04 | 11.09 | 1.856 | 1.21 |
| **15MPa** | **N-BEATS** | **100** | **0.0099** | **24** | **12** | **3.98** | **4.06** | **0.465** | **0.381** |
| | TCN | 400 | 0.838 | 24 | 12 | 5.46 | 5.49 | 0.645 | 0.535 |
| | RNN | 400 | 0.188 | 24 | 12 | 28.12 | 25.68 | 4.491 | 2.39 |
| | TF | 400 | 0.327 | 24 | 12 | 25.52 | 23.43 | 3.954 | 2.98 |
| **25MPa** | N-BEATS | 100 | 0.0291 | 100 | 20 | 4.62 | 4.54 | 0.957 | 0.961 |
| | **TCN** | **400** | **0.575** | **24** | **12** | **2.88** | **2.85** | **0.704** | **0.592** |
| | RNN | 400 | 0.872 | 24 | 12 | 12.33 | 10.05 | 1.950 | 1.21 |
| | TF | 400 | 0.398 | 24 | 12 | 10.05 | 11.73 | 8.232 | 2.45 |
| **35MPa** | N-BEATS | 100 | 0.0277 | 50 | 10 | 4.26 | 4.34 | 0.640 | 0.473 |
| | **TCN** | **400** | **0.348** | **24** | **12** | **4.46** | **4.34** | **0.540** | **0.472** |
| | RNN | 400 | 0.233 | 24 | 12 | 22.18 | 24.98 | 2.983 | 1.79 |
| | TF | 400 | 0.568 | 24 | 12 | 15.08 | 13.06 | 2.221 | 2.13 |

**The outcome from the statistical baseline models**

The Theta model consistently excels in forecasting the axial strain responses. It achieved the lowest statistical error metrics across the datasets, with RMSE values ranging from 0.53 to 0.85, MAE from 0.47 to 0.80, MAPE from 2.18% to 7.31%, and SMAPE from 2.18% to 7.61% (Table 6). Notably, at a confining pressure of 5 MPa, the Theta model recorded with MAPE of 2.42%, SMAPE of 2.42%, RMSE of 0.55, and MAE of 0.48, as reflected in Fig. 12.

The ES model also demonstrated strong performance at specific confining pressure levels, achieving the lowest error metrics at 25 MPa and maintaining good accuracy at 35 MPa. The TBATS model effectively handled complex seasonality at 15 MPa but did not consistently outperform the Theta model. While AutoARIMA provided reliable forecasts, it was generally less effective than the Theta and deep learning models, as indicated by its higher evaluation metrics (Table 6). The results clearly illustrate that the Theta model is the most effective choice for accurate predictions across varying confining pressure levels.

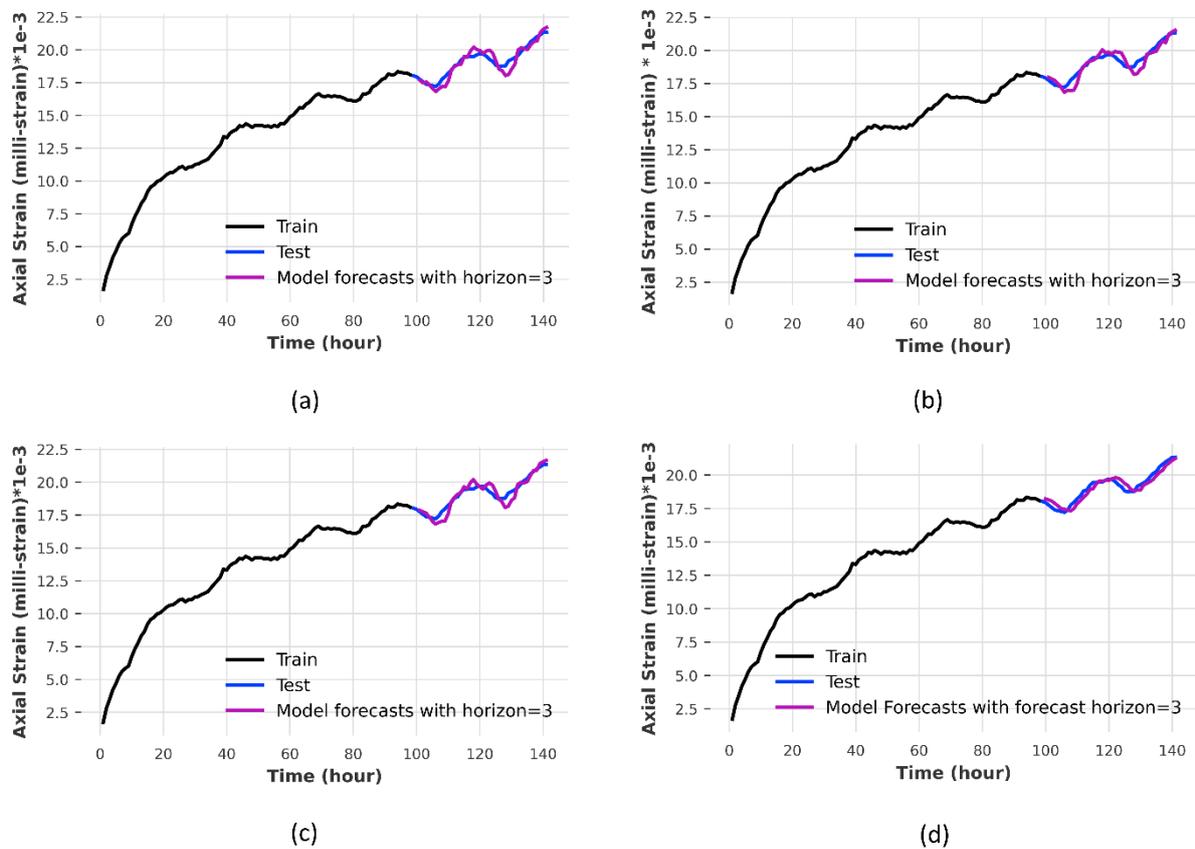

(a)  (b)  (c)  (d)

**Fig. 12:** An example from statistical baseline models such as for (a) ES, (b) TBATS, (c) AutoARIMA, and (d) Theta, respectively, at a confining pressure level of 5MPa. In comparison, the Theta model outperformed for minimal evaluation metrics error values.

**Table 6:** Summary of statistical evaluation metrics for statistical baseline models, where bold highlighted for the best models

| Sr No | TS datasets | Classical or Baseline Models | MAPE (%) | SMAPE (%) | RMSE | MAE |
|---|---|---|---|---|---|---|
| 1 | 5MPa | ES | 16.04 | 14.49 | 3.72 | 3.19 |
|   |   | TBATS | 17.22 | 15.48 | 3.95 | 3.45 |
|   |   | Auto ARIMA | 17.36 | 15.59 | 3.98 | 3.45 |
|   |   | **Theta** | **2.42** | **2.42** | **0.55** | **0.48** |
| 2 | 15MPa | ES | 7.56 | 7.85 | 0.92 | 0.77 |
|   |   | TBATS | 5.25 | 5.16 | 0.59 | 0.51 |
|   |   | Auto ARIMA | 5.82 | 5.88 | 0.68 | 0.58 |
|   |   | **Theta** | **6.35** | **6.06** | **0.74** | **0.60** |
| 3 | 25MPa | ES | 32.55 | 26.90 | 8.16 | 7.06 |
|   |   | TBATS | 2.66 | 2.65 | 0.72 | 0.57 |
|   |   | Auto ARIMA | 8.30 | 7.88 | 2.04 | 1.79 |
|   |   | **Theta** | **2.18** | **2.18** | **0.53** | **0.47** |

| 4 | 35MPa | ES | 2.93 | 2.99 | 0.40 | 0.32 |
|---|---|---|---|---|---|---|
| | | TBATS | 20.04 | 22.66 | 2.39 | 2.21 |
| | | Auto ARIMA | 4.85 | 4.71 | 0.68 | 0.53 |
| | | **Theta** | **7.31** | **7.61** | **0.85** | **0.80** |

**Key outcome from analytical models**

We employed notable analytical models to validate and support the deep learning models, including the Kelvin, Burgers, power law, spring-dashpot, and logarithmic models, with predicted values illustrated in Fig. 13 across various pressure levels. The power law model demonstrated the best fit with $R^2$ (0.98), yielding creep parameters C (0.10 day$^{-1}$) and k (0.40) (Tables 7, and 8). The Burgers model followed closely with fitting parameters A (0.10), B (0.022 × 10$^{14}$ s$^{-1}$), and k (2.64 day$^{-1}$) and an $R^2$ of 0.97. Evaluation metrics in Table 7a revealed superior forecasting performance for the power law model, exhibiting RMSE (0.33), MAE (0.28), MAPE (1.45%), and SAMPE (1.46%), compared to the Burgers model, which had RMSE (0.52), MAE (0.44), MAPE (3.799%), and SAMPE (3.98%).

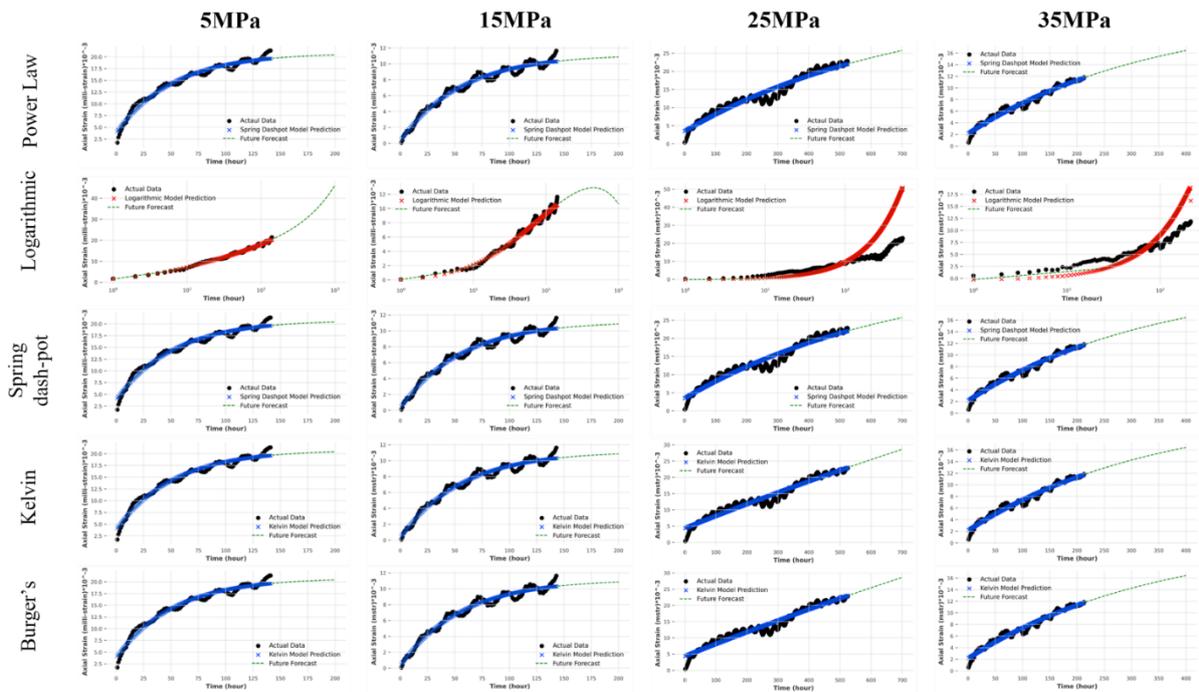

**Fig. 13:** Analytical models for long-term prediction, such as Burger's, Kelvin, spring-dashpot, power law, and logarithmic, for creeping results of axial strain (milli-strain) over time (hour), forecast trends from left to right at different confining pressure level conditions.

**Table 7:** Performance metrics of the analytical models from training data (best results are highlighted)

| Sr. No. | data | Analytical Models | $R^2$ | RMSE | MAE | MAPE (%) | SMAPE (%) |
|---|---|---|---|---|---|---|---|
| 1 | 5MPa | **Power Law** | **0.98** | **0.33** | **0.28** | **1.45** | **1.46** |
| | | Burger | 0.97 | 0.52 | 0.42 | 3.49 | 3.55 |
| | | Kelvin | 0.95 | 0.92 | 0.91 | 6.05 | 5.95 |
| | | Spring dashpot | 0.93 | 0.93 | 0.91 | 6.70 | 6.51 |
| | | Logarithmic | 0.92 | 1.01 | 0.97 | 6.91 | 6.72 |
| 2 | 15MPa | **Power Law** | **0.97** | **0.36** | **0.34** | **2.46** | **2.42** |
| | | Burger | 0.97 | 0.52 | 0.42 | 3.49 | 3.55 |
| | | Kelvin | 0.94 | 0.92 | 0.91 | 6.05 | 5.95 |
| | | Spring dashpot | 0.93 | 0.93 | 0.91 | 6.70 | 6.51 |
| | | Logarithmic | 0.92 | 1.01 | 0.97 | 6.91 | 6.72 |
| 3 | 25MPa | **Power Law** | **0.98** | **0.36** | **0.29** | **1.49** | **1.46** |
| | | Burger | 0.96 | 0.52 | 0.42 | 3.49 | 3.55 |
| | | Kelvin | 0.95 | 0.92 | 0.91 | 6.05 | 5.95 |
| | | Spring dashpot | 0.92 | 0.93 | 0.91 | 6.70 | 6.51 |
| | | Logarithmic | 0.91 | 1.01 | 0.97 | 6.91 | 6.72 |
| 4 | 35MPa | **Power Law** | **0.97** | **0.42** | **0.38** | **2.53** | **2.48** |
| | | Burger | 0.94 | 0.52 | 0.42 | 3.49 | 3.55 |
| | | Kelvin | 0.93 | 0.92 | 0.91 | 6.05 | 5.95 |
| | | Spring dashpot | 0.93 | 0.93 | 0.91 | 6.70 | 6.51 |
| | | Logarithmic | 0.90 | 1.01 | 0.97 | 6.91 | 6.72 |

**Table 8:** Creep parameters obtained by fitting the best two considered analytical models, such as Power-law creep followed by Burger Creep, recorded for the homogeneous rock salt facies (Temperature: 27°C = 300K).

| MSTL creep data stages | Confining pressure (MPa) | Differential Axial stress (MPa) | Total axial stress (MPa) | Creep duration (Days/hour sample) | Power law creep | | Burger creep | | |
|---|---|---|---|---|---|---|---|---|---|
| | | | | | C (day$^{-n}$) | k (-) | $B_1$ (-) | $B_2$ (*1.16.10$^4$s$^{1)}$) | $k_2$ (day$^{-1}$) |
| Stage 1 | 5 | 8 | 13 | 5.8/141 | 0.10 | 0.44 | 0.10 | 0.022 | 2.64 |
| Stage 2 | 15 | 8.2 | 23.2 | 5.8/143 | 0.04 | 0.58 | 0.05 | 0.011 | 1.09 |
| Stage 3 (in-situ) | 25 | 8.4 | 33.4 | 22/520 | 0.04 | 0.58 | 0.07 | 0.008 | 0.57 |
| Stage 4 | 35 | 8.6 | 43.6 | 9/216 | 0.04 | 0.53 | 0.03 | 0.010 | 2.40 |

**Discussions**

This study assesses the long-term creep deformation at ambient temperature under a range of confining pressures. Minimal causality between axial strain and temperature was confirmed by STL analysis, which broke down the creep datasets into trend, seasonal, and residual components. The data's linearity was evaluated using significant causality tests, such as Granger and ADF (Table 4). For time-dependent deformation, DNN models—N-BEATS, TCN, RNN, and Transformers in particular—showed excellent predictive ability, with low RMSE, MAE, MAPE, and SMAPE values (Table 5). Traditional baseline models, such as AuToARIMA and ES, on the other hand, fared better under different conditions (Table 6). The incorporation of mathematical models strengthened the DNN results' resilience. The study demonstrates how well sophisticated DNN models work with the Darts library to forecast time series. With preliminary data analysis, the investigation verified that the triaxial creep data is stationary, non-seasonal, and increases linearly with time, demonstrating the data's dependability for upcoming projections. Axial strain and temperature are not causally related, emphasizing the need for more sophisticated modeling methods to produce reliable forecasts. These important forecasting model performances were discussed in detail as summarized below;

**Forecasting performance of DNN models**

The N-BEATS model forecasts both the trend and seasonal components by decomposing down time series data into trend, seasonal, and residual components. This allows it to achieve exceptional precision and interpretability. likewise, the TCN model forecasts the residual term, which allows for reliable long-term strain response forecasting outside of test datasets. Precise decision-making enabled both the N-BEATS and TCN models to achieve high forecast accuracy; these models' respective RMSE values are 3.7441 for N-BEATS and 4.5044 for TCN, and their MAPE values are 2.50% and 2.001% (as seen in Table 5). Traditional statistical approaches, such as the Theta model and ES, by contrast, increased average RMSE by 20.18%, MAE by 17.5%, and MAPE by 15% under different confining pressures (as seen in Table 6). The TCN's robustness for enormous time series datasets is increased by its architecture, which includes residual blocks and causal convolution. N-BEATS's great accuracy and efficacy are further confirmed by its effective training. When it comes to forecasting, N-BEATS and TCN routinely perform better than other DNN models.

**Forecasting performance of statistical models**

The Theta model demonstrated superior performance based on the analysis of statistical baseline models—ES, Auto ARIMA, Theta, and TBATS. Across the board, lower error metrics, including RMSE, MAE, MAPE (%), and SMAPE (%), support this conclusion. The Theta model consistently minimized prediction errors, as evidenced by the lowest RMSE, and provided the most reliable forecasts, indicated by the smallest MAE. Furthermore, its robust handling of relative error, with the lowest MAPE, and its balanced performance in managing both overestimations and underestimations, reflected in the superior SMAPE, underscore its accuracy and reliability in forecasting axial strain and time over the given forecast horizon of 3. These findings establish the Theta model as the most suitable and preferred method among the models evaluated, particularly for time series forecasting tasks where precision and consistency are critical.

**Comparative significance analysis and advantages of DNN over analytical models**

Deep Neural Network models—particularly N-BEATS and Temporal Convolutional Networks—demonstrate significantly higher predictive performance metrics by approximately 15–20%, compared to traditional analytical models such as the Power Law and Burger models in predicting axial creep strain. This improvement is supported by superior evaluation metrics including $R^2$, RMSE, MAE, MAPE, and SMAPE. While analytical models like the Power Law ($R^2$ = 0.97–0.98, RMSE = 0.33–0.42) and Burger ($R^2$ = 0.94-0.97, RMSE = 0.52) are effective in fitting static or short-term trends, they are limited in capturing the dynamic temporal dependencies inherent in long-term creep behavior. In contrast, DNN models dynamically learn from data through techniques such as window-shifting and data-splitting, enabling them to produce long-term forecasts with data-driven validation. This capability is crucial for accurately modeling non-linear and time-dependent phenomena. Analytical models are primarily curve-fitting approaches that describe the general trend of the data. Although the Power Law and Burger models performed relatively well among the analytical approaches, outperforming other models like Kelvin's and spring-dashpot models, as shown in Tables 7 and 8, they still fall short in accurately representing complex temporal patterns. Their inability to model unseen future response limits their reliability for long-term prediction when observational data is unavailable.

On the other hand, DNN models such as N-BEATS and TCN are inherently more adaptive. They effectively capture intricate temporal dependencies and non-linear variations in the data, making them particularly suitable for dynamic systems. By segmenting datasets into training and testing windows, these models allow for controlled validation against actual observations of axial creep strain over extended periods. This approach not only enhances predictive precision but also ensures model reliability by tuning data lengths to maintain consistency and accuracy.

Thus, DNN models are far more capable than traditional analytical models for applications requiring precise long-term forecasting. Their dynamic architecture and ability to learn from time-dependent data make them the preferred choice for modeling real-world time-dependent deformation response in complex systems.

**Statistical significance of the results: MCB test**

We next compared the predictions produced by the N-BEATS and TCN models to those produced by other benchmark forecasters to ascertain the statistical significance of the predictions. RMSE, MAE, MAPE (%), and SMAPE (%) scores across different time series datasets and confining pressure level conditions at their corresponding critical distances were used to calculate the average ranks of the various methods to assess their relative performance using multiple comparisons with the best test. According to the MCB test findings in Table 9 and Figs. 14–18, the N-BEATS model performed better across the board in all measures at the confining pressure level of 5–35 MPa, consistently achieving the lowest average 1. For the best-fitted model, the reference value is the upper bound of the critical distance. Critical intervals for all benchmark forecasters (in terms of RMSE, MAE, MAPE, and SMAPE scores) were higher than the reference value for DNN models, suggesting noticeably poorer performance compared to N-BEATS. Other baseline forecasters had non-overlapping intervals, indicating their poorer performance, but N-BEATS and TCN showed a minor overlap for the SMAPE metric. The N-BEATS and TCN models regularly outperformed the others, as seen by the MCB plots, which also showed that the theta model had the lowest mean rank among statistical baseline models.

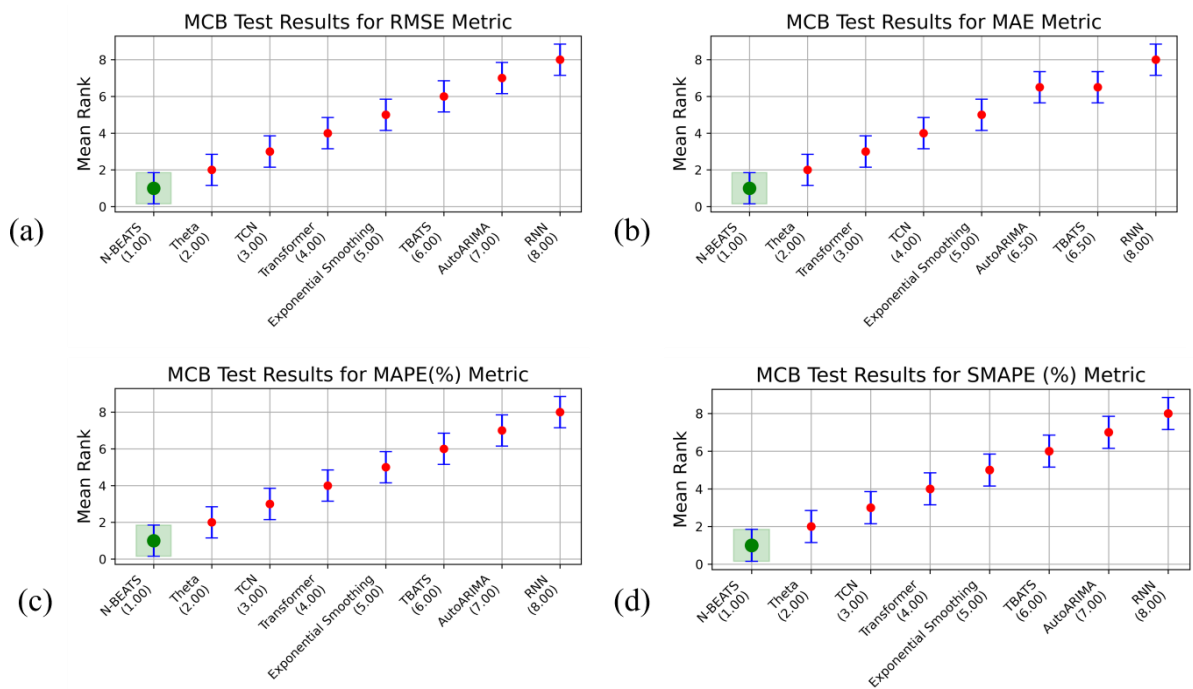

**Fig. 14:** Visualization of Multiple Comparisons with the Best test analysis with mean rank value across four performance metrics: (a) RMSE, (b) MAE, (c) MAPE in %, and (d) SMAPE in % at a confining pressure level of 5MPa, respectively. Each subplot presents the mean rank of different forecasting models, where lower ranks signify better performance. For instance, in subplot (a), 'N-BEATS-1' indicates that the N-BEATS model achieved the top rank for RMSE. This comprehensive visualization offers insights into model performance across various evaluation criteria, similar to others.

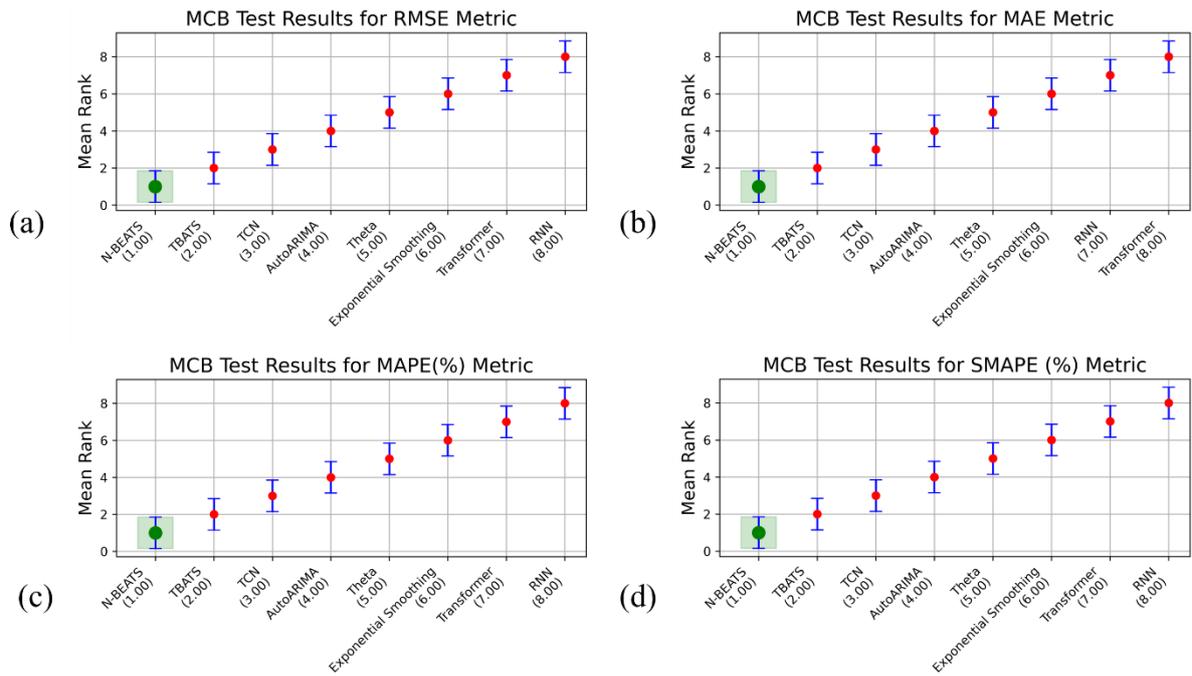

**Fig. 15:** Visualization of Multiple Comparisons with the Best test analysis with mean rank value across four performance metrics: (a) RMSE, (b) MAE, (c) MAPE in %, and (d) SMAPE in % at 15MPa, respectively.

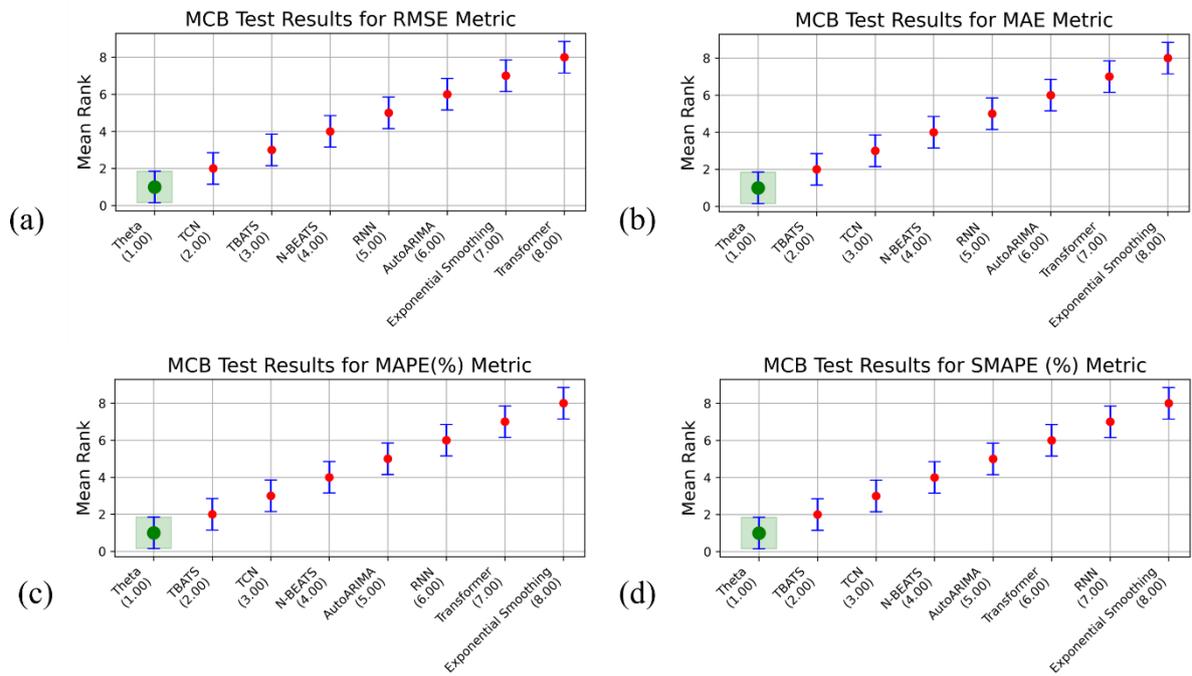

**Fig. 16:** Visualization of MCB test analysis with mean rank value across four performance metrics: (a) RMSE, (b) MAE, (c) MAPE in %, and (d) SMAPE in % at 25MPa, respectively.

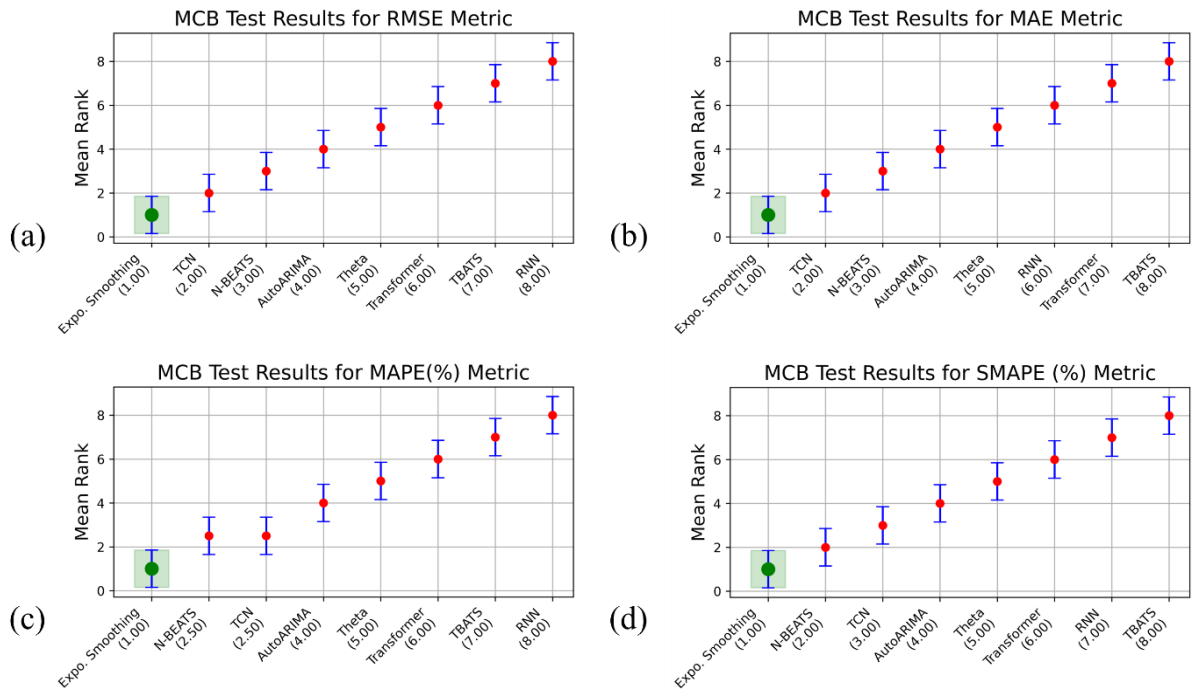

**Fig. 17:** Visualization of MCB test analysis with mean rank across four performance metrics: (a) RMSE, (b) MAE, (c) MAPE in %, and (d) SMAPE in % at 35MPa, respectively.

**Table 9:** Average rank derived from the MCB test for DNN and statistical baseline models corresponding to the performance metrics (the best one is highlighted).

| Datasets | Models | Mean Rank | | | |
|---|---|---|---|---|---|
| | | RMSE | MAE | MAPE (%) | SMAPE (%) |
| 5MPa | **N-BEATS** | 1 | 1 | 1 | 1 |
| | **TCN** | 3 | 4 | 3 | 3 |
| | RNN | 8 | 8 | 8 | 8 |
| | Transformer | 4 | 3 | 4 | 4 |
| | ES | 5 | 5 | 5 | 5 |
| | TBATS | 6 | 6.5 | 6 | 6 |
| | Auto ARIMA | 7 | 6.5 | 7 | 7 |
| | **Theta** | 2 | 2 | 2 | 2 |
| 15MPa | **N-BEATS** | 1 | 1 | 1 | 1 |
| | **TCN** | 3 | 3 | 3 | 3 |
| | RNN | 8 | 7 | 8 | 7 |
| | Transformer | 7 | 8 | 7 | 7 |
| | ES | 6 | 6 | 6 | 6 |
| | **TBATS** | 2 | 2 | 2 | 2 |
| | Auto ARIMA | 4 | 4 | 4 | 4 |
| | Theta | 5 | 5 | 5 | 5 |
| **25MPa** | **N-BEATS** | 4 | 4 | 4 | 4 |
| | **TCN** | 3 | 3 | 3 | 3 |

|       | Model      | RMSE | MAE | MAPE | SMAPE |
|-------|------------|------|-----|------|-------|
|       | RNN        | 5    | 5   | 6    | 7     |
|       | Transformer| 8    | 7   | 7    | 6     |
|       | ES         | 7    | 8   | 8    | 8     |
|       | **TBATS**  | 3    | 2   | 2    | 2     |
|       | Auto ARIMA | 6    | 6   | 5    | 5     |
|       | **Theta**  | 1    | 1   | 1    | 1     |
| **35MPa** | N-BEATS | 3    | 3   | 2.5  | 2     |
|       | **TCN**    | 2    | 2   | 2.5  | 3     |
|       | RNN        | 8    | 6   | 8    | 6     |
|       | Transformer| 6    | 7   | 6    | 6     |
|       | **ES**     | 1    | 1   | 1    | 1     |
|       | TBATS      | 7    | 8   | 7    | 8     |
|       | Auto ARIMA | 4    | 4   | 4    | 4     |
|       | Theta      | 5    | 5   | 5    | 5     |

Fig. 18, a heatmap (left) and error analysis (right) for the MCB test in another statistical way is used to visualize and test for improved performance predictive models in terms of statistical evaluation error analysis metrics. This was also done for confining pressure at 5MPa and creep data of axial strain responses. To find the best model for each metric, such as RMSE, MAE, SMAPE, and SMAPE, look for the models with the reddest cell. The best model is indicated in red, and its rank number is displayed in bold black in the error plot, which displays the average rank of models over metrics with an 85% confidence interval length.

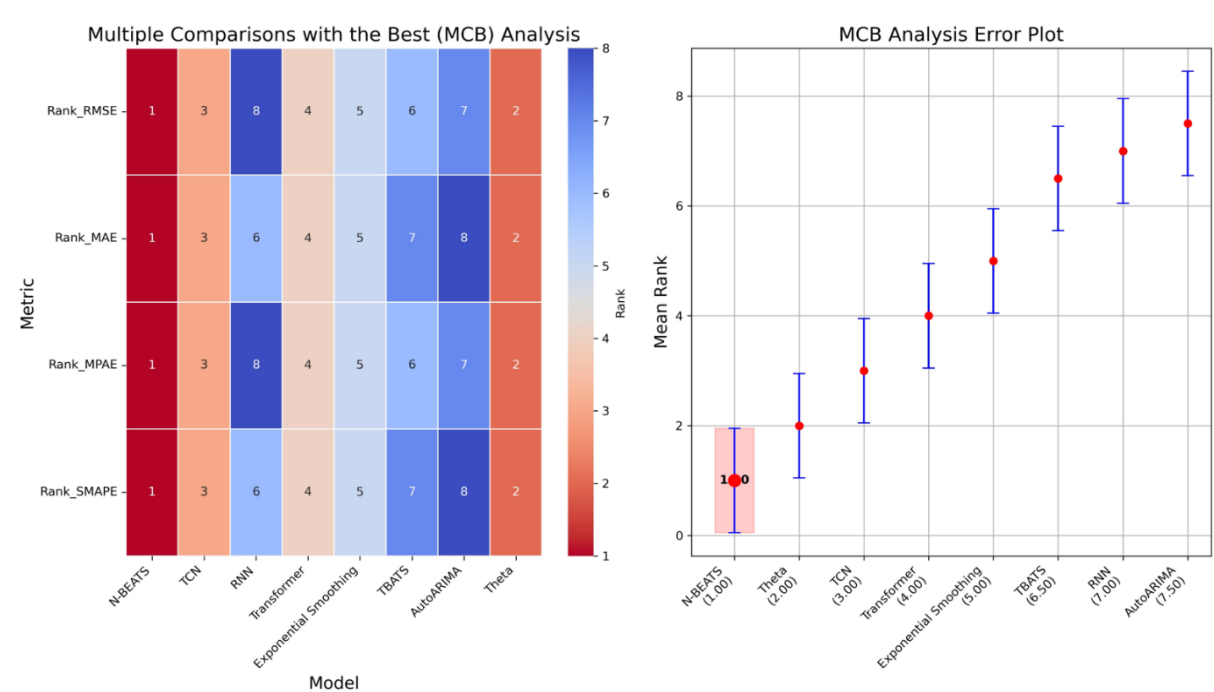

**Fig. 18:** Another example of a box view for MCB analysis (left) with error plot (right) for DNN and statistical baseline models for metrics such as RMSE, MAE, MAPE (%), and SMAPE (%) at a confining

pressure level of 5MPa, respectively. The mean/average rank value of the N-BEATS model is 1, which shows the best model in terms of accuracy among other models.

**Conformal prediction for N-BEATS and TCN models**

After analyzing the MCB test and mean ranks to facilitate the best DNN models for forecasting for time series datasets at different confining pressure level conditions, we introduced two best DNN models for time series datasets with superior evaluation metrics that promise to forecast, we conducted the conformal prediction with 95% prediction intervals between actual and model predicted. The N-BEATS overperformed, followed by the TCN model, with evaluation metrics such as RMSE, MAE, MAPE, and SMAPE (Table 4) in terms of accuracy compared to other DNN models. That is why we focus on CP with the two best proposed DNN models in this case for analysis, shown in Figs. 19-20, respectively. CP provides a probabilistic way of generating prediction intervals, which is crucial for industrial applications where point forecasts are seldom needed [32].

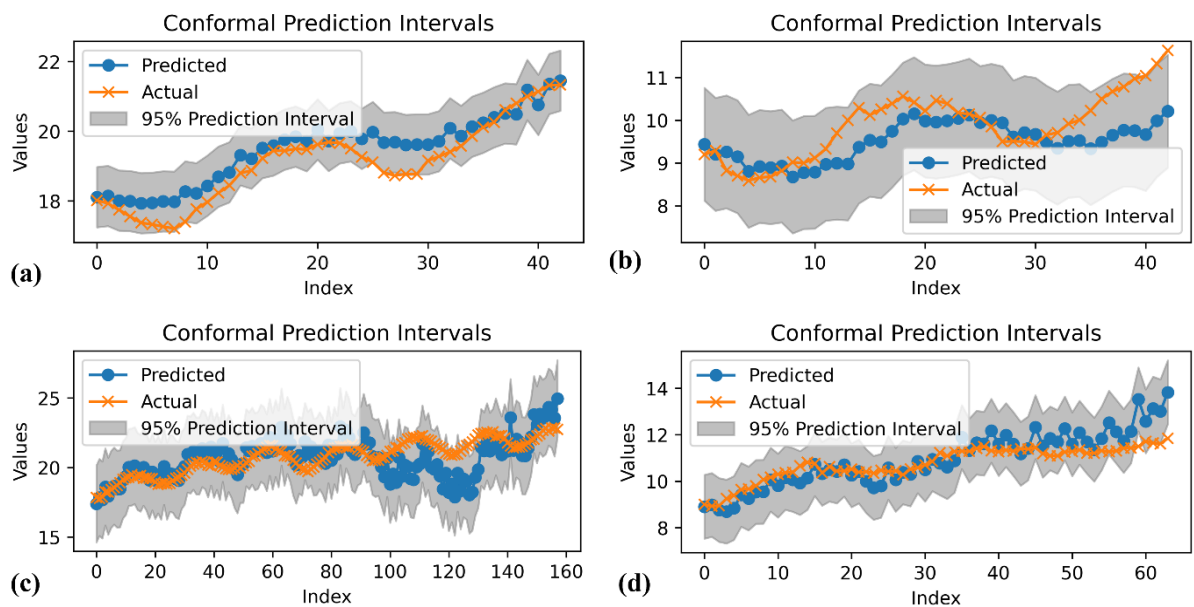

**Fig. 19:** The proposed N-BEATS utilized for conformal prediction between precedent against actual data in 95% prediction intervals for (a) 5MPa, (b) 15MPa, (c) 25MPa, and (d) 35MPa, respectively. The horizontal axis corresponds to the index (time in hour), and the vertical axis to axial strain values (milli-strain×$10^{-3}$)

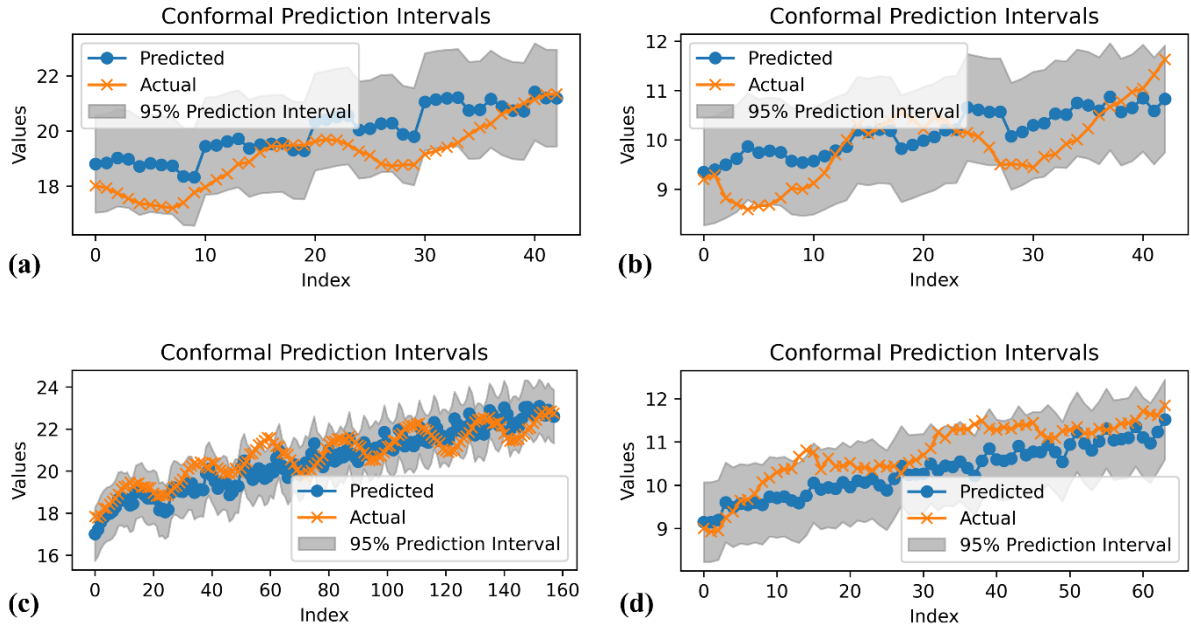

**Fig. 20:** The TCN model followed by the N-BEATS was utilized for conformal prediction between precedent against actual data in 95% prediction intervals for (a) 5MPa, (b) 15MPa, (c) 25MPa, and (d) 35MPa, respectively. The horizontal axis corresponds to the index (time in hours), and the vertical axis corresponds to axial strain values (milli-strain×$10^{-3}$).

**Model forecasts: N-BEATS and TCN**

The Darts Python library has revolutionized time series forecasting with its user-friendly interface and diverse range of models, from classical approaches to cutting-edge deep neural networks. This library simplifies the forecasting process by employing familiar methods such as .fit() and .predict(), akin to scikit-learn, enabling seamless model training and forecasting. One of Darts' standout features is its robust support for univariate and multivariate time series, accommodating potentially large datasets with multiple series. Moreover, the library facilitates backtesting, model combination, and integration of external data, enhancing the accuracy and flexibility of predictions and forecasting. Notably, Darts includes ML-based models capable of probabilistic forecasting, providing valuable insights into forecast uncertainty. Models like N-BEATS and TCN have demonstrated remarkable performance beyond sample datasets.

In our evaluations, the N-BEATS model excelled in short-term forecasting, efficiently capturing patterns in time series data with shorter durations. Conversely, the TCN model proved superior in forecasting over longer durations, demonstrating its effectiveness in capturing complex trends and patterns. Fig. 21 illustrates the N-BEATS model successfully forecasting 100-hour samples beyond the set, highlighting its robust performance in the short term. Meanwhile, Fig. 22 showcases the TCN model accurately forecasting up to 200-hour samples, underscoring its capability in longer-term forecasts.

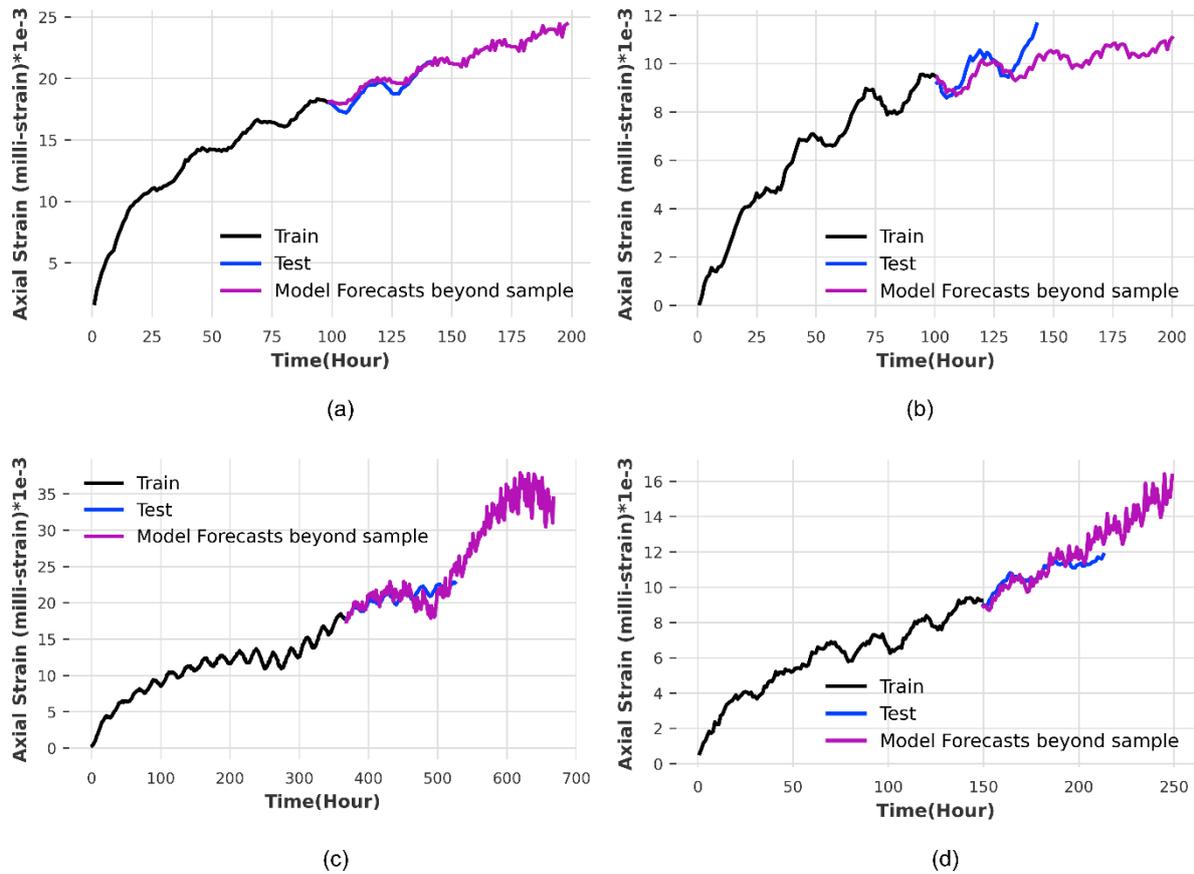

**Fig. 21:** The N-BEATS model forecasts beyond the test sample for various confining pressure conditions, such as for (a) 5MPa, (15) 15MPa, (c) 25MPa, and (d) 35MPa, respectively. The N-BEATS model forecasts that the next value is realistic, limited to the length of the test duration.

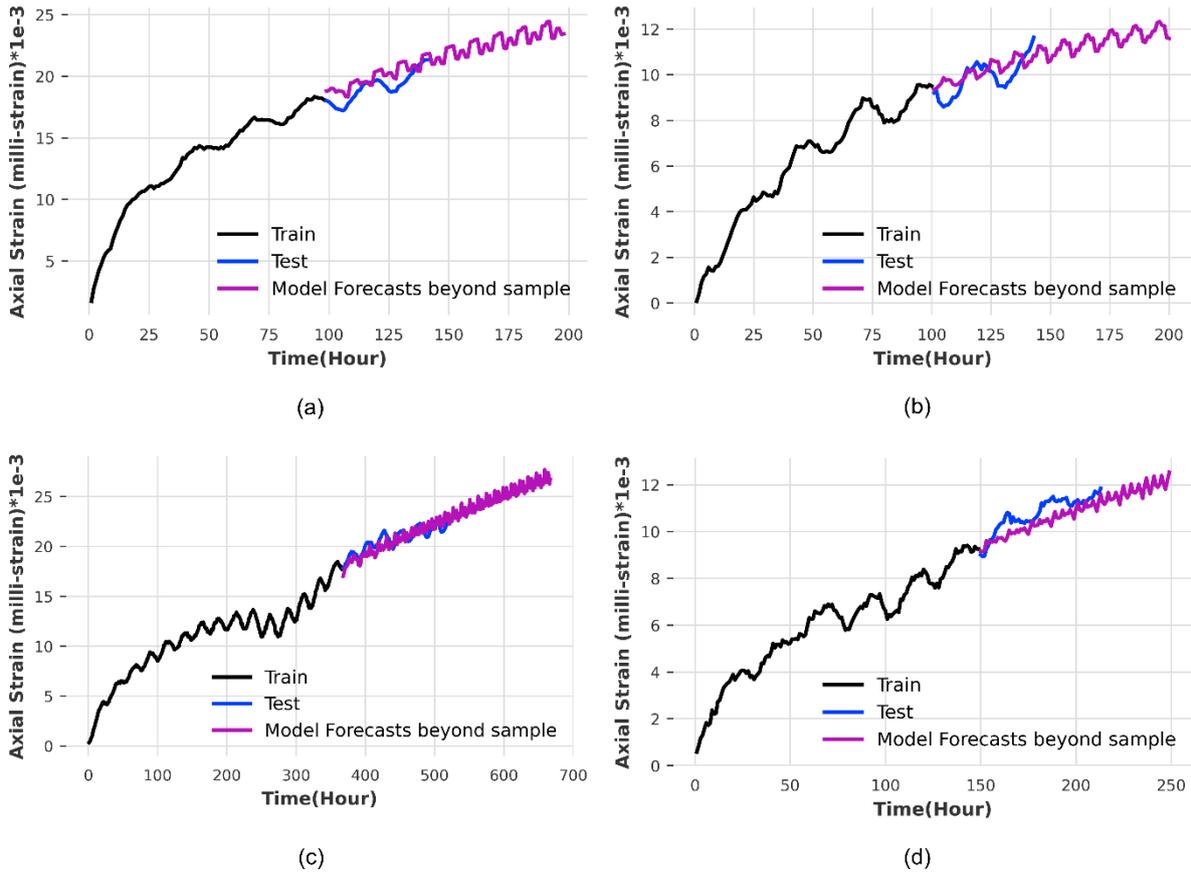

**Fig. 22:** The TCN model forecasts beyond the test sample for various confining pressure conditions, such as for (a) 5MPa, (15) 15MPa, (c) 25MPa, and (d) 35MPa, respectively. The TCN model forecasts the next value to be realistic, limited to the length of the test duration.

**Implications and limitations**

The study underscores the limitations of traditional causality tests and analytical models in forecasting axial strain deformation response over extended periods. Specifically, DNN models—such as N-BEATS and TCN—demonstrate significant advantages in capturing complex temporal dependencies and patterns, yielding more accurate forecasts in dynamic systems. Their applicability extends beyond energy storage to other critical fields, including weather forecasting and geoscience, where precise forecasting is essential.

While DNN models show strong forecasting performance, they also have certain limitations, such as high complexity and substantial computational resource requirements, particularly for GPU acceleration. Additionally, they may struggle with data spikes, leading to prediction inaccuracies, especially when data is limited. Nevertheless, DNNs have key advantages in forecasting creep deformation, as they capture temporal dependencies and allow for trend validation over both short and long-term scenarios. In contrast, analytical models typically fit data trends over time but cannot validate forecasts without additional future data, which limits their reliability for long-term trend validation. Unlike DNNs, analytical models do not require a defined forecast window length; they simply fit data over time to obtain a trend or pattern. For example, using a five-day dataset of axial strain creep deformation, an analytical model might provide a trend over an extended period. However, without future data, these models cannot validate or challenge the forecasted trend, posing a significant drawback compared to DNNs. In DNN models, however, a time-shifting procedure for input and output windows can be applied, allowing for the division of creep datasets into training and testing sets. This enables model validation against actual recorded data, supporting both forecasting and validation of axial strain deformation trends. Consequently, DNN models play a pivotal role in addressing the limitations of traditional methods for long-term creep deformation forecasting.

Future work could enhance DNN forecasting accuracy by incorporating temperature fluctuations and other engineering features, further improving the model performance of rock formation's creep response under in situ conditions. Overall, these findings support a shift towards advanced DNN models for more robust and accurate long-term forecasting.

**Conclusions**

This study highlights the exceptional performance of deep neural network (DNN) models in forecasting long-term axial strain under varying confining pressures at room temperature.

In this study, some important outcomes of this paper are as below;

- STL decomposition and causality tests revealed no significant relationship between strain and temperature, with minimal seasonality observed. The stationary p-value (< 0.05) and identifiable patterns in creep deformation suggest the necessity of advanced modeling.
- The N-BEATS and TCN models led in predictive accuracy with notably low on-average RMSE (3.7441, 4.5044) and MAPE in % (2.50, 2.001) respectively, outperforming other DNNs (RNN, Transformer) and statistical models (e.g., ES, Theta), demonstrating high precision in time series forecasting.
- N-BEATS and TCN showed a 15-20% performance metrics improvement over traditional models, capturing intricate temporal patterns and dependencies, which analytical models, limited to linear trends, could not achieve.
- The study establishes a foundation for applying DNNs to time-dependent deformation modeling. Future work could expand these approaches to other forecasting challenges in the field of geophysics, enhancing the understanding of complex patterns in similar domains.

**Acknowledgments**

The authors extend their sincere gratitude to the Commonwealth Scientific and Industrial Research Organization (CSIRO), Australia, for providing data support and invaluable insights that were instrumental in completing this research. We also acknowledge the Research and Development (R&D) division at the Indian Institute of Technology (ISM) Dhanbad, Jharkhand, for their support in permitting this research for publication. Additionally, the authors gratefully acknowledge funding from Sorbonne University, which enabled the completion and publication of this work.

**AUTHORS DECLARATION**

**Credit Authorship Contribution Statement**

Pradeep Kumar Shukla: Methodology, deep learning techniques, Coding script, Formal analysis, Writing – original draft. Tanujit Chakraborty: Investigation, conceptualization, formal analysis, writing – review, editing, and funding acquisition. Mustafa Sari: Data curation, Formal analysis, writing review, and editing. Joel Sarout: Data curation, writing review, and editing. Partha Pratim Mandal: Conceptualization, Methodology, Validation, Supervision, Writing – review & editing.

**Declaration of Competing Interest**

The authors declare that they have no known competing financial interests or personal relationships that could have appeared to influence the work reported in this paper.

**Conflict of Interest**

The author declares on behalf of all authors in this paper that no conflict of interest exists in submitting this manuscript and that all authors approve the publication of the manuscript.

**Additional information**

Correspondence and requests for material should be addressed to P.K.S.

**References**

1. Hou B, Shangguan S, Niu Y, et al. Unique properties of rock salt and application of salt caverns on underground energy storage: a mini review. *Energy Sources, Part A: Recovery, Utilization and Environmental Effects*. 2024;46(1):621-635. doi:10.1080/15567036.2023.2288295

2. Ramesh Kumar K, Hajibeygi H. Multiscale simulation of inelastic creep deformation for geological rocks. *J Comput Phys*. 2021;440. doi:10.1016/j.jcp.2021.110439

3. Ramesh Kumar K, Makhmutov A, Spiers CJ, Hajibeygi H. Geomechanical simulation of energy storage in salt formations. *Sci Rep*. 2021;11(1):1-24. doi:10.1038/s41598-021-99161-8

4. Sarout J, Sari M, Esteban L, et al. Short-term deformation and yield, long-term creep, and sealing capacity of a Devonian rock salt at 1 km depth in the Canning Basin (WA). 2023;(September).

5. Gao R, Wu F, Chen J, Zhu C, Ji C. Study on creep characteristics and constitutive model of typical argillaceous salt rock in energy storage caverns in China. *J Energy Storage*. 2022;50(January):104248. doi:10.1016/j.est.2022.104248

6. Zhao K, Ma H, Zhou J, et al. Rock Salt Under Cyclic Loading with High-Stress Intervals. *Rock Mech Rock Eng*. 2022;55(7):4031-4049. doi:10.1007/s00603-022-02848-1

7. Yang Y, Zoback M. Viscoplastic Deformation of the Bakken and Adjacent Formations and Its Relation to Hydraulic Fracture Growth. *Rock Mech Rock Eng*. 2016;49(2):689-698. doi:10.1007/s00603-015-0866-z

8. Mallants D, Bourdet J, Camilleri M, et al. An Assessment of Deep Borehole Disposal Post-Closure Safety. *Nucl Technol*. 2023;210(9):1511-1534. doi:10.1080/00295450.2023.2266609

9. Mallants D, Beiraghdar Y, Doblin C, et al. Deep borehole disposal of intermediate-level waste: progress from Australia's RD&D project. *Proceedings of the INMM & ESARDA Joint Virtual Annual Meeting August 23-26 & August 30-September 1*. 2021;(August):1-14.

10. Yang C, Daemen JJK, Yin JH. Experimental investigation of creep behavior of salt rock. *International Journal of Rock Mechanics and Mining Sciences*. 1999;36(2):233-242. doi:10.1016/S0148-9062(98)00187-9

11. Mandal PP, Sarout J, Rezaee R. Geomechanical appraisal and prospectivity analysis of the Goldwyer shale accounting for stress variation and formation anisotropy. *International Journal of Rock Mechanics and Mining Sciences*. 2020;135(September). doi:10.1016/j.ijrmms.2020.104513

12. Mandal PP, Sarout J, Rezaee R. Triaxial Deformation of the Goldwyer Gas Shale at In Situ Stress Conditions—Part I: Anisotropy of Elastic and Mechanical Properties. *Rock Mech Rock Eng*. 2022;55(10):6121-6149. doi:10.1007/s00603-022-02936-2

13. Lyu C, Liu J, Zhao C, Ren Y, Liang C. Creep-damage constitutive model based on fractional derivatives and its application in salt cavern gas storage. *J Energy Storage*. 2021;44. doi:10.1016/j.est.2021.103403

14. Wu F, Gao R, Zou Q, Chen J, Liu W, Peng K. Long-term strength determination and nonlinear creep damage constitutive model of salt rock based on multistage creep test: Implications for underground natural gas storage in salt cavern. *Energy Sci Eng*. 2020;8(5):1592-1603. doi:10.1002/ese3.617

15. Gao R, Wu F, Chen J, Zhu C, Ji C. Study on creep characteristics and constitutive model of typical argillaceous salt rock in energy storage caverns in China. *J Energy Storage*. 2022;50. doi:10.1016/j.est.2022.104248

16. Xu T, Zhou G, Heap MJ, Yang S, Konietzky H, Baud P. The Modeling of Time-Dependent Deformation and Fracturing of Brittle Rocks Under Varying Confining and Pore Pressures. *Rock Mech Rock Eng*. 2018;51(10):3241-3263. doi:10.1007/s00603-018-1491-4

17. Mcnee F, Gorter J, Glass F. New Age Dating of Evaporites in Canning Basin, WA, Australia. A Case Study based on samples from the Frome Rocks Salt Diapir New Age Dating of Evaporites in Canning Basin , WA , Australia . A Case Study based on samples from the Frome Rocks Salt Diapir. *3rd Australian Exploration Geoscience Conference*. 2021;(July).

18. Erő J, Szűcs J. Statistical Model. *Nuclear Structure Study with Neutrons*. Published online 1974:477-482. doi:10.1007/978-1-4613-4499-5-24

19. Zhang B, Song C, Jiang X, Li Y. Electricity price forecast based on the STL-TCN-NBEATS model. *Heliyon*. 2023;9(1):e13029. doi:10.1016/j.heliyon.2023.e13029

20. Sohrabbeig A, Ardakanian O, Musilek P. Decompose and Conquer: Time Series Forecasting with Multiseasonal Trend Decomposition Using Loess. *Forecasting*. 2023;5(4):684-696. doi:10.3390/forecast5040037

21. Panja M, Chakraborty T, Kumar U, Liu N. Epicasting: An Ensemble Wavelet Neural Network for forecasting epidemics. *Neural Networks*. 2023;165:185-212. doi:10.1016/j.neunet.2023.05.049

22. Oreshkin BN, Carpov D, Chapados N, Bengio Y. N-Beats: Neural Basis Expansion Analysis for Interpretable Time Series Forecasting. *8th International Conference on Learning Representations, ICLR 2020*. Published online 2020:1-31.

23. Rob J. Hyndman, Yeasmin Khandakar. Automatic Time Series Forecasting: The forecast Package for R. *J Stat Softw*. 2008;27(3):22. https://doi.org/10.18637/jss.v027.i03

24. Lim B, Zohren S. Time-series forecasting with deep learning: A survey. *Philosophical Transactions of the Royal Society A: Mathematical, Physical and Engineering Sciences*. 2021;379(2194). doi:10.1098/rsta.2020.0209

25. Pandya V, Follow ·. Use Tensorflow LSTM for Time Series Forecasting Time Series data. Published online 2023:1-21.

26. de Livera AM, Hyndman RJ, Snyder RD. Forecasting time series with complex seasonal patterns using exponential smoothing. *J Am Stat Assoc*. 2011;106(496):1513-1527. doi:10.1198/jasa.2011.tm09771

27. Lara-Benítez P, Carranza-García M, Riquelme JC. An Experimental Review on Deep Learning Architectures for Time Series Forecasting. *Int J Neural Syst*. 2021;31(3). doi:10.1142/S0129065721300011

28. Zhang Q, Song Z, Wang J, Zhang Y, Wang T. Creep Properties and Constitutive Model of Salt Rock. *Advances in Civil Engineering*. 2021;2021. doi:10.1155/2021/8867673

29. Assimakopoulos V, Nikolopoulos K. The theta model: A decomposition approach to forecasting. *Int J Forecast*. 2000;16(4):521-530. doi:10.1016/S0169-2070(00)00066-2

30. Agunloye OK, Shangodoyin DK, Arnab R. Lag length specification in Engle-Granger cointegration test: a modified Koyck mean lag approach based on partial correlation. *Statistics in Transition new series*. 2014;15(4):559-572. doi:10.59170/stattrans-2014-037

31. Bai S, Kolter JZ, Koltun V. An Empirical Evaluation of Generic Convolutional and Recurrent Networks for Sequence Modeling. Published online 2018.

32. Wu P, Sun J, Chang X, et al. Data-driven reduced order model with temporal convolutional neural network. *Comput Methods Appl Mech Eng*. 2020;360. doi:10.1016/j.cma.2019.112766

33. Cho K, Van Merriënboer B, Gulcehre C, et al. Learning phrase representations using RNN encoder-decoder for statistical machine translation. *EMNLP 2014 - 2014 Conference on Empirical Methods in Natural Language Processing, Proceedings of the Conference*. Published online 2014:1724-1734. doi:10.3115/v1/d14-1179

34. Vaswani A, Shazeer N, Parmar N, et al. Attention is all you need. *Adv Neural Inf Process Syst*. 2017;2017-Decem(Nips):5999-6009.

35. Liu HZ, Xie HQ, He JD, Xiao ML, Zhuo L. Nonlinear creep damage constitutive model for soft rocks. *Mech Time Depend Mater*. 2017;21(1):73-96. doi:10.1007/s11043-016-9319-7

36. Makhmutov A. Nonlinear 2D Finite Element Modeling: Cyclic Energy Storage in Salt Caverns with Creep Deformation Physics. 2020;(August).

37. Park BY, Sobolik SR, Herrick CG. Geomechanical Model Calibration Using Field Measurements for a Petroleum Reserve. *Rock Mech Rock Eng*. 2018;51(3):925-943. doi:10.1007/s00603-017-1370-4

38. Sone H, Zoback MD. Time-dependent deformation of shale gas reservoir rocks and its long-term effect on the in situ state of stress. *International Journal of Rock Mechanics and Mining Sciences*. 2014;69:120-132. doi:10.1016/j.ijrmms.2014.04.002

39. Mandal PP, Sarout J, Rezaee R. *Triaxial Deformation of the Goldwyer Gas Shale at In Situ Stress Conditions—Part II: Viscoelastic Creep/Relaxation and Frictional Failure*. Vol 56. Springer Vienna; 2023. doi:10.1007/s00603-023-03437-6

40. Gong, F., Xu, L., Gao, M., Zhao, Y. & Zhang, P. Compressive damage constitutive model for brittle coal based on the compaction effect and linear energy dissipation law. *Int J Coal Sci Technol* 12, 43 (2025).

41. Chen, Y., Hao, X., Xue, D., Li, Z. & Ma, X. Creep behavior and permeability evolution of coal pillar dam for underground water reservoir. *Int J Coal Sci Technol* 10, (2023).

42. Yin, J. J., Lei, J., Fan, K. & Wang, S. Integrating image processing and deep learning for effective analysis and classification of dust pollution in mining processes. *Int J Coal Sci Technol* 10, (2023).

43. Jiang, Y. *et al.* Mechanical properties and acoustic emission characteristics of soft rock with different water contents under dynamic disturbance. *Int J Coal Sci Technol* 11, (2024).

44. Yuan, R. *et al.* Time effect of elastic energy release of surrounding rock and evaluation method of reasonable advancing speed. *Int J Coal Sci Technol* 12, (2025).

45. Yan, F., Wang, E., Liu, X., Qi, C. & Jia, W. Experimental study on strain localization and slow deformation evolution in small-scale specimens. *Int J Coal Sci Technol* 12, (2025).

46. Huang, F. *et al.* Slope stability prediction based on a long short-term memory neural network: comparisons with convolutional neural networks, support vector machines and random forest models. *Int J Coal Sci Technol* 10, (2023).